\documentclass[12pt]{article}

\usepackage{graphicx}
\usepackage{slashed}
\usepackage{amsmath,amssymb}
\usepackage{bm}
\usepackage{epsfig}

\newcommand
{\bpsi}{\bar{\psi}}

\begin{document}

\title{ \bf
Low energy constituent  quark and pion effective  couplings 
 in a weak  external magnetic field
}
 
\author{ F\'abio L. Braghin
\\
{\normalsize Instituto de F\'\i sica, Federal University of Goias,}
\\
{\normalsize Av. Esperan\c ca, s/n,
 74690-900, Goi\^ania, GO, Brazil}
}

\maketitle 

\abstract{
An effective model  with  pions and constituent quarks
 in the presence of a weak external background 
electromagnetic field is derived  
by  starting  from a dressed one gluon exchange 
quark-quark  interaction.
By applying the auxiliary field and background field methods,
the structureless pion limit is considered to extract effective pion and
constituent  quark couplings in the presence of a weak magnetic field.
The leading terms of a  
large quark and gluon masses
 expansion are obtained by resolving
effective coupling constants which turn out to depend on a weak magnetic field.
Two pion field definitions are considered for that.
Several relations between the effective coupling constants and parameters 
can be derived exactly or in the limit of very large quark mass at zero and weak 
constant magnetic field. 
Among these ratios,  the
Gell Mann Oakes Renner and the quark level Goldberger Treiman relations are obtained.
In addition to that, in the pion sector,  the leading terms of Chiral Perturbation 
Theory coupled to the electromagnetic field  is recovered.
 Some numerical estimates  are provided
 for the effective coupling constants and parameters.
}



\section{Introduction}
\label{intro}

The intrincated structure of Quantum Chromodynamics (QCD)
makes very difficult the identification  of the relation between 
its  fundamental degrees of freedom  and the models that describe
quite well hadron and nuclear Physics in the vacuum and in a certain range of variables such as
 temperature, baryonic density, magnetic field and so on.
Low energy effective models are usually based 
on phenomenology and also general theoretical 
results and symmetries from QCD.
Effective field theory (EFT)   have been developed and strengthened
since the formulation of  Chiral Perturbation Theory (ChPT) 
\cite{cpt1,cpt2,cpt4,weinberg-1979}
and they   contribute for establishing
 these conceptual and calculational gaps between the two
 levels in the description of
strong interactions systems.
However they do not provide microscopic 
first ground numerical predictions for 
 the low energy 
coefficients.
First principles calculations provided by lattice QCD 
also contributes to these programs since they  should offer
the ultimate full theoretical numerical   predictions.
However it is also very  interesting   to establish sound relations  between 
 low energy effective models and  EFT 
 with  QCD by deriving them from QCD.
 Different  mechanisms have already been found
to yield a large variety of effective couplings between hadrons, 
in particular quark-antiquark light mesons chiral dynamics have been 
investigated extensively with focus on reproducing the low energy coefficients
of chiral perturbation theory  
with progressively improved  account of QCD non Abelian effects,
for example in Refs.
 \cite{simonov,osipov-etal,ERV,PRC,holdom,wang-etal,NJL,EPJA-2016}.
Quark-based  models such as NJL  \cite{NJL}
 and constituent quark models
 \cite{derafael,lavelle-mcm,thomas1}
are some of the most successful ones  describing a large variety of
mesons and baryons  observables 
and properties with 
least number of  parameters.
One of the most important ingredients  of these models emerges
from  both  the phenomenology 
of hadrons   and  also QCD: 
approximate chiral symmetry and its dynamical symmetry breaking (DChSB)
\cite{SWbook}. 
The relevance of the chiral condensate as the order parameter for the DChSB 
is widely recognized 
in spite of some  recent controverse about  its location
\cite{in-hadron-1,in-hadron-2,reinhardt-weigel}.
In this context, it was claimed that the chiral constituent quark model 
\cite{derafael,manohar-georgi} can
give rise to
a large Nc  EFT that copes  the large Nc expansion and 
the constituent quark model 
in Ref. \cite{weinberg-2010}. 
 This EFT is composed by the leading large $N_c$ terms for 
constituent quarks coupled to pions and constituent gluons, 
besides the 
leading terms of chiral perturbation theory.
This EFT has   been derived recently 
  from a single  leading term  of the QCD effective action by considering
quark polarization for 
a  dressed (non perturbative) gluon exchange between quarks
 \cite{EPJA-2016}.
 In the present work this derivation is addressed again with the coupling 
to a background electromagnetic field
 by considering one term
of the QCD effective action that is the 
dressed one gluon exchange between quarks as discussed below.

Hadron dynamics 
 in the presence of 
 magnetic fields ($B_0$) have attracted the attention due to sizeable values
in several systems such as in 
magnetars, 
$B_0 \simeq 10^{15}$G, and non central relativistic heavy ions collisions (RHIC),
 $B_0 \simeq 10^{18}$G, in spite of being formed for very short time interval
  \cite{tuchin}. 
Even such strong magnetic fields can be considered to be relatively  small
with respect to some hadron energy scales.
For example, one may have a  large constituent
quark effective mass with respect to  
a weak magnetic 
 field limit,  and
a small  parameter  such as
$eB_0/{M^*}^2$ can be considered.
For a  constituent quark
effective mass ${M^*} \simeq 0.31$GeV and $eB_0/{M^*}^2 \sim 0.6$
it follows 
$eB_0 \sim 0.06$GeV$^2$
  or equivalently  $eB_0 
\sim 3 m_\pi^2$.
By considering further that,  with the magnetic catalysis
 the quark effective mass increases with the 
magnetic field, it might reach $M^* \sim 0.45$GeV
for  $B_0 \sim 7 m_{\pi}^2$, expected to 
 appear in RHIC,
the  parameter of the expansion is  
$eB_0/{M^*}^2 \sim 0.6$.
This parameter is completely compatible with a weak electromagnetic 
background field such that: $|e A_\mu| << M^*$ or  $|e A_\mu| < M^*$.
  Many phenomena are expected to  take place under  finite magnetic field 
\cite{reviews-B-qcd,review-B-general,magn-catalysis,Meff-B,chiral-cond,magnetic-inibition,mag-inibition-2,CP-vio-B,chernodub,ch-imbal-1,cme-1,ch-asym-1,ch-imb-invmagcat}.
In particular  it has been found  that 
a magnetic field can yield modifications and corrections
in the strength and type of the hadron interactions, in particular
constituent 
quarks, gluons  and pions 
\cite{G-B-2,krein-etal,ayala-g2,G-B-3,alpha-B,PRD-2016,PRD-2017c}.

The following low energy
 quark effective global color model \cite{ERV,PRC}
coupled to a background electromagnetic field
  will be considered: 
\begin{eqnarray} \label{Seff}  
Z = N \int {\cal D}[\bpsi, \psi]
\exp \;  i \int_x  \left[
\bar{\psi} \left( i \slashed{D} 
- m \right) \psi 
-
 \frac{g^2}{2}\int_y j_{\mu}^b (x) 
{\tilde{R}}^{\mu \nu}_{bc} (x-y) j_{\nu}^{c} (y) 
+ \bpsi J + J
^* \psi \right]  ,
\end{eqnarray}
where the color  quark current is 
$j^{\mu}_a = \bar{\psi} \lambda_a \gamma^{\mu} \psi$, 
 the sums in color, flavor and Dirac indices are implicit, $\int_x$ 
stands for 
$\int d^4 x$,
$a,b...=1,...(N_c^2-1)$ stand 
for color in the adjoint representation, 
and $i,j,k=0,1..$ will be used  for SU(2) flavor indices,
and the covariant quark derivative  with the minimal coupling to the
background photon
 is:
$D_{\mu} = \partial_\mu  - i e Q  A_{\mu}$
with  the diagonal matrix 
$\hat{Q} =  diag(2/3, -1/3)$.
In several  gauges the gluon kernel
can  be  written in terms of the transversal and longitudinal components,
$R_T(k)$ and $R_L(k)$,
as:
\begin{eqnarray}
\tilde{R}^{\mu\nu}_{ab} (k) = \delta_{ab} \left[
 \left(g^{\mu\nu} - \frac{k^\mu k^\nu}{k^2}
\right)  R_T (k) 
+  \frac{k^\mu  k^\nu}{ k^2}
R_L (k) \right].
\end{eqnarray}
Even if other terms   arise from the 
non Abelian structure of the gluon sector,
 the quark-quark  
interaction  
   (\ref{Seff})  
is a leading term of the QCD effective action.
The use of a dressed (non perturbative) gluon propagator 
already takes into account non abelian contributions that
garantees
  important effects.
This approximation for the QCD effective action 
also is basically equivalent to the Schwinger Dyson  
equations at the rainbown ladder approximation \cite{SD-rainbow}.
Furthermore, it will be assumed and required
 that this dressed  gluon propagator provides enough strength for 
DChSB
as obtained for example in
 \cite{SD-rainbow,kondo,cornwall,maris-craig,higa,aoki,gluonmass}. 
Although it is possible to include the contribution
of further 
terms from the QCD effective action \cite{wang-etal}
the aim of this work is to identify and to compute 
 the resulting pion and constituent quark  interactions 
resulting from a one loop background field  analysis 
for the interaction (\ref{Seff}).
Along this work the same method  proposed in
Ref. \cite{EPJA-2016} will be considered with the use of the 
background field method to introduce the sea and constituent quarks
and of the auxilary field method to introduce the light quark-antiquark 
chiral states and mesons.
Vector currents associated to light vector mesons are neglected since they 
represent
 heavier states and mesons being    less important at lower energies.
Two different  non linear definitions of the pion field
are chosen  by performing  chiral rotations and they are exhibited  in the Appendix
A.
Nevertheless all the steps of the calculation 
 will be presented  to 
incorporate the coupling to background  photons. 
The resulting sea quark determinant is expanded for  large quark and gluon 
 masses.

 Interactions of pions with constituent quarks 
within constituent quark models are  expected to be 
 equivalent to pion-nucleons couplings and our results 
reinforce this view.
Firstly, effective couplings are written in terms of an external  photon field 
that,  in a second step,
is  reduced to a 
weak magnetic field $B_0$  that is  incorporated into the effective coupling constants.
Several exact or approximated  ratios of the 
effective coupling constants and parameters are calculated.
Among them, the Gell Mann Oakes Renner \cite{GOR} and 
the Goldberger Treiman  \cite{GT-ori,GT} relations arise at zero or weak magnetic fields.
 Numerical estimates for the resulting effective coupling constants
and parameters
are  presented in Section (5).
In spite of the fact that the gluon propagator and the quark-gluon vertex
may
present magnetic field dependence as well  these will not be considered.
Since the gluon propagator is not really completely known
two  different gluon propagators
are considered to provide numerical estimates.
 This work therefore extends the work presented in 
\cite{EPJA-2016}
by including the electromagnetic coupling of pions and constituent quarks.
Some leading  corrections to pion and constituent quark
effective coupling constants due to weak magnetic field 
are also calculated  in the logics of Ref. \cite{PRD-2016}
where exclusively constituent quark effective self  interactions had
been considered.
The pion sector has been   investigated in similar and even more 
general calculations by considering their structure and by  taking into account
 gluon 3-point and 4-point Green's functions 
\cite{ERV,PRC,wang-etal,E-R1986}.
The pion sector  will be  presented as a sort of benchmark to 
assure the procedure, within the limit of structureless mesons,
 reproduces all the leading terms of  chiral perturbation theory even if
the resulting expressions for the low energy coefficients
 do not contain all the structures from a 
more general calculation. 
The work  is organized as follows.
The background field method and the auxiliary field method
are briefly reminded in Section (\ref{sec:two-Q}).
The two pion field definitions are presented shortly in the Appendix A and 
the
 equivalence between the leading  pion and constituent couplings 
to the background photon  of the 
two pion field definitions 
is shown in Appendix B.
In section (\ref{sec:expansion-W}),
 the quark determinant is expanded in the leading 
order by choosing the pion field in the Weinberg's definition in terms of 
covariant derivatives \cite{W-pionfield} and the effective
 coupling constants are resolved.
Ratios between effective coupling constants are exhibited.
In Section (\ref{sec:expansion-U}), the usual 
pion field in terms of exponential representation,
$U, U^\dagger$, is 
considered for the expansion of the corresponding quark determinant.
It yields, in the pion sector, Chiral Perturbation Theory coupled to the 
electromagnetic field. 
The constituent quark sector in the leading order 
yields  the most relevant pion-quark effective couplings
that are also  calculated in the presence of the external electromagnetic field.
Relations between the effective coupling constants are also exhibited
at zero and weak magnetic field and numerical estimates are presented
in Section (\ref{sec:numerics}).
In the last Section there is a Summary.

\section{ Flavor structure and light mesons fields }
\label{sec:two-Q}

A Fierz transformation 
\cite{ERV,PRC,NJL}  is performed 
for the interaction (\ref{Seff}) 
and  the color singlet terms are selected as usually.
{  
These non singlet color terms that are already 
smaller than the color singlet 
 by a factor 
$1/N_c$.
It will be shown just before Section (\ref{sub:gap})
they yield  only higher orders interaction terms
being  outside of the scope of the work.
}
The Fierz transformation of the 
 quark interaction above, denoted here by ${\cal F} (\Omega) = \Omega_F$, 
 can be rewritten
in terms of 
 bilocal quark currents,
$j_i^q(x,y) =  \bpsi (x) \Gamma^q \psi (y)$ where 
$q=s,p,v,a$
and  
$\Gamma_{s} = I_{2} . I_4$ (for the 2x2 flavor and 4x4  Dirac  identities),
 $\Gamma_{p} = i  \gamma_5 \sigma_i$,
$\Gamma_{v}^\mu =  \gamma^\mu \sigma_i $ and
$\Gamma_{a}^\mu =    i \gamma_5 \gamma^\mu  \sigma_i $,
where   $\sigma_i$ are the flavor SU(2) Pauli matrices.
The resulting non local  interactions $\Omega_F$ are the following:
\begin{eqnarray} \label{fierz4}
 \Omega &\equiv& 
g^2 j_{\mu}^a (x) \tilde{R}^{\mu\nu}_{ab} (x-y) j_{\nu}^b (y)
\\
\Omega_F &=&
4 \alpha g^2 
\left\{ 
\left[ j_S (x,y)  j_S(y,x) + j_P^i(x,y)  j_P^i(y,x)  \right] R  (x-y)
\right.
\nonumber
\\
&-& \left.
  \frac{1}{2} \left[ j_{\mu}^i (x,y) j_{\nu}^i (y,x) 
+  {j_{\mu}^i}_A (x,y)  {j_{\nu}^i}_A (y,x)
\right]  \bar{R}^{\mu\nu} (x-y) \right\},
\nonumber
\end{eqnarray}
where 
$i,j,k=0,...(N_f^2-1)$ and 
$4 \alpha =  8/9$ for  SU(2)  flavor.
The kernels above
 can  be written as:
\begin{eqnarray} \label{Rk}
R (x-y) &\equiv& 
3 R_T (x-y) + R_L (x-y),
\nonumber
\\ \label{Rbar}
\bar{R}^{\mu\nu} (x-y) &\equiv& 
g^{\mu\nu} (R_T (x-y) +R_L (x-y) )
+ 2 \frac{\partial^{\mu} \partial^{\nu}}{\partial^2} 
(R_T (x-y)  - R_L (x-y) ).
\end{eqnarray}
A longwavelength  local limit of expression (\ref{fierz4})
 yields the Nambu Jona Lasinio (NJL) and vector NJL couplings.

 By employing the background field method (BFM) 
\cite{BFM,SWbook}
 it will be considered that
quarks might either  form light mesons and the chiral condensate
 or to be constituent quarks for baryons.
So the quark field  will be splitted into  the
(constituent quark)  background field ($\psi_1$)
 and the sea quark field ($\psi_2$) that  will be integrated out.
For this, the following shift is performed: 
 $\psi \to \psi_1 + \psi_2$.
This sort of decomposition is not exclusive to the 
BFM and it is found in other approaches   \cite{PRD-2001}.
{ 
The field $\psi_2, \bpsi_2$ will be integrated out
and,  for this,
the action in (\ref{Seff}) 
is expanded 
for  the lowest one loop order as:
\begin{eqnarray} \label{exp-BFM}
S \simeq  S_{eff} [\bpsi_1, \psi_1]+ 
\int  \left( {J}^* + \frac{\delta S_{eff} [\bpsi_1, \psi_1]}{\delta \psi_1} \right) \psi_2 + 
 \int \bpsi_2 \left( \frac{\delta S_{eff} [\bpsi_1, \psi_1]}{\delta \bpsi_1} + J  \right) 
+  \int \bpsi_2 \frac{\delta S_{eff} [\bpsi_1, \psi_1]}{\delta \psi_1 \delta \bpsi_1} \psi_2
+ ... ,
\end{eqnarray}
where the first term is the tree level contribution 
and the linear terms  are simply  zero.
}
Therefore at the one loop BFM 
 level only  the quadratic term contributes and
it yield the same result as  
   to perform the field splitting 
for the  bilinears 
 $\bpsi \Gamma_q \psi$ \cite{BFM}.
It  can be written that:
\begin{eqnarray} \label{split-Q} 
\bpsi \Gamma^q \psi &\to& (\bpsi \Gamma^q \psi)_2 + (\bpsi \Gamma^q \psi)_1,
\end{eqnarray}
where  $(\bpsi \psi)_2$ 
will be treated in the usual  way as valence quark of the NJL model.
After a rearrangment the resulting full interaction $\Omega_F$ is splitted accordingly
$$
\Omega_F  \to \Omega_1 + \Omega_2 + \Omega_{12} ,
$$
where $\Omega_{12}$ contains the interactions between the two components.
This separation 
   preserves chiral symmetry,
and it might  not be  simply  a  mode separation  of low and high energies.
Besides that, the above shift of quark bilinears
looks   suitable for incorporating  quark-anti-quark
states which are the most relevant states for the  low energy 
 below the nucleon mass scale.
To make possible to integrate out the component $\psi_2$,
the interactions $\Omega_2$
can  be handled in  two ways:
\\
(i)  By considering   a weak field approximation it can be neglected
 $\Omega_2 < < \Omega_1$.
This yields directly the one loop BFM that might receive corrections by 
a perturbative expansion which incorporates $\Omega_2$ \cite{BFM};
\\
(ii)  By resorting to the 
auxiliary field method (AFM) with the introduction of 
a set of auxiliary fields  by means of unitary functional integrals
in  the generating functional
 \cite{ERV,PRC,EPJA-2016,kleinert}.
A further advantadge of doing so is that the auxiliary fields (a.f.)
 allow to incorporate properly  DChSB with the formation of the 
scalar quark condensate which endows quarks with a large effective mass.
Therefore the second alternative   improves  the one loop
BFM.
Bilocal auxiliary fields ($
S(x,y), P_i(x,y), V_\mu^i(x,y)$ and $\bar{A}_\mu^i(x,y)$)
 will be  introduced by multiplying the generating functional 
by the following normalized Gaussian integrals:
\begin{eqnarray}  \label{af}
 1 &=& N \int D[S] D[P_i]
 e^{- \frac{i}{2 }  
\int_{x,y}   R  \alpha  \left[ (S - g   j^S_{(2)})^2 +
(P_i -  g    j^{P}_{i,(2)} )^2 \right]}
 \int 
D[V_\mu^i]
 e^{- \frac{i}{4 } 
\int_{x,y} {\bar{R}^{\mu\nu}} \alpha
\left[ (V^i_{\mu} -  g    j_{V,\mu}^{i,(2)}) (V^i_{\nu} -
 g   j_{V,\nu} ^{i,(2)} )
\right] }
\nonumber
\\
&&
\int  D[\bar{A}_\mu^i] 
 e^{- \frac{i}{4 }  
\int_{x,y} {\bar{R}^{\mu\nu}} \alpha
\left[ 
  ( \bar{A}^i_{\mu} -  g    {j_{A,\mu}^{ i,(2)}} ) ( \bar{A}^i_{\nu} -
 g   {j_{A,\nu}^{i,(2)}} )
\right]
}.
\end{eqnarray}
The bilocal a.f. in this expression
 have been shifted by quark currents such as to cancel out the fourth 
order interactions $\Omega_2$.
These  shifts have unit Jacobian 
 and they generate a.f.  couplings to valence  quarks.
 The procedure for the auxiliary field adopted in the present work
for the meson sector
follows  the development proposed in Refs.
\cite{PRC,ERV,E-R1986}.
This  is discussed next by considering however the 
structureless mesons limit.
Besides the meson sector however it contains bilinears of 
constituent quark field. 
They will compose
interaction terms for baryons and mesons.
The bilocal auxiliary fields  can reduce to punctual  meson fields
by expanding in an infinite basis of local meson  fields \cite{PRC}, for instance 
a particular bilocal field 
$\phi_\alpha(x,y) = S(x,y), P_i(x,y),V_\mu^i (x,y),\bar{A}_\mu^i(x,y)$ 
can be writtten in terms of a corresponding
 complete orhonormal
sum of local fields  $\xi_\alpha^k (u)$ as:
\begin{eqnarray}
\phi_\alpha (x,y) = \Phi_\alpha \left( \frac{x+y}{2}, x-y \right)
= \Phi_\alpha (u ,z) = {F}_\alpha (z)  
+ \sum_k F_{k,\alpha} (z) \xi_\alpha^k (u),
\end{eqnarray}
where $F_\alpha, F_{k,\alpha}$ are vacuum functions invariant under translation for
 each of the bilocal mode and corresponding local field $\xi_\alpha^k (z)$
The low energy regime amounts to
 picking up only the low lying (lighest) modes $k=0$
and making the form factors  
to reduce to the zero momentum  limit
$F_{k,\alpha}(z) = F_{k,\alpha} (0)$, i.e. to constants.
At the end the very  long-wavelength limit
can be reached by selecting the leading terms of these expansion
which roughly correspond to
  simply considering
 the local limit for structureless lightest  mesons.
In this case $\xi_\alpha (u) = \xi_\alpha (x)$.
This work is concerned with   weak magnetic field effects  in the low energy regime,
the leading local field states of  composite fields can be considered 
also  the  effects of magnetic field on the meson structure will be neglected.
As a matter of fact, these effects should be relevant for much 
 larger magnetic fields.
Moreover, in the low energy regime  the  vector mesons should not be relevant
to the dynamics since they are considerably heavier than the pion.
In this limit, the above valence quark coupling to the local meson fields
from expression (\ref{af}),
by omitting the index $_2$,
reduces to:
\begin{eqnarray} \label{scalars-q}
\bpsi (x) \Xi (x,y) \psi (y)   \simeq 
\bpsi (x)  F \left[ 
 s (x) + p_i (x) \gamma_5 \sigma_i  
  \right]
 \delta (x-y)
\psi (y) ,
\end{eqnarray}
where $F$ is the pion decay constant that allows for the canonical definition of the pion field
as $\pi_i = F p_i$.

By considering the identity
$\det A = \exp \; Tr \; \ln (A)$,
the  Gaussian integration of the valence quark field
yields:
\begin{eqnarray} \label{Seff-det}  
I_{det}   &=&   \; Tr  \; \ln \; \left\{
i  S^{-1} (x-y) \right\}
\end{eqnarray}
where
\begin{eqnarray} \label{Sq-1}
S^{-1} (x-y) &=&
 {S_{0,c}}^{-1} (x-y) 
+ \Xi (x-y)
\\
&+&  
 2   R (x-y) \alpha g^2
 \left[  (\bpsi (y) \psi(x))
+ i  \gamma_5 \sigma_i  (\bpsi (y) i \gamma_5  \sigma_i \psi (x)) \right]
\nonumber
\\
&-& 
\alpha g^2 \bar{R}^{\mu\nu} (x-y) \gamma_\mu  \sigma_i \left[
 \bpsi (y) \gamma_\nu  \sigma_i \psi(x)
+  i \gamma_5  (\bpsi  (y)
i \gamma_5 \gamma_\nu  \sigma_i \psi (x) \right]
,
\nonumber
\end{eqnarray}
where 
$Tr$ stands for traces of discrete internal indices 
and integration of  spacetime coordinates,
the quark kernel was defined as
$S_{0,c}^{-1} (x-y) = \left(  i \slashed{D} -  m
\right) \delta (x-y)$.
In the absence of the external photon field 
$S_{0}^{-1} (x-y) = \left(  i \slashed{\partial} -  m
\right) \delta (x-y)$.
 The above determinant has been already  investigated in different limits.
By neglecting the constituent quark field, it arises 
a low energy model in terms of low lying meson states  
 \cite{PRC,E-R1986,thooft}.
By considering only the electromagnetic field, without mesons and quarks,
an Euler-Heisenberg-type  effective model  emerges \cite{EH1936,cond-B,schwinger51}.
By considering only  the  constituent quark currents  
higher order quark effective interactions
were obtained \cite{PLB2016}. 
It was possible to 
trace back the symmetry breaking effective interactions to
the quark mass and curiously these effective interactions were found
to have strengths of the order of the chiral invariant effective interactions.
Besides that, the conclusions from  large Nc  Witten analysis \cite{witten} were verified
since the couplings behave as $N_c^{1-n}$ for $n$ quarks interactions.
Finally pions and constituent quark currents have been considered for zero magnetic field
in \cite{EPJA-2016} and the resulting model turns out to 
be the large Nc effective field theory (EFT) proposed by Weinberg \cite{weinberg-2010}.

{ 
Now it is   shown that the color non singlet terms neglected
in expression (\ref{fierz4}) yield  higher order contributions.
Consider  terms such as 
$1/N_c R (x-y)(\bpsi \lambda_\alpha \psi) (\bpsi \lambda_\alpha \psi)$
 where $\lambda_\alpha$
are the Gell Mann color matrices.
In the one loop BFM method their contribution in $\Omega_2$
are  simply neglected \cite{BFM}.
The contribution of this term appears only in the
integration of the quark field in expression (\ref{Sq-1})
with extra terms of the type:
$\Delta S^{-1} (x-y) = \lambda_\alpha
R (x-y)  g^2  \bpsi \lambda _\alpha  \psi + ...$
When performing the expansion of the determinant  these terms 
only contribute in colorless combinations with extra factor $1/N_c$ for 
each quark current.
 Therefore they do not  contribute to  the leading pion and constituent 
quark sectors investigated in this work.
If 
colorfull auxiliary fields $\tilde{S}_\alpha$ were  introduced 
to make possible the integration of the corresponding term in $\Omega_2$,
these fields  would be not observables  and they would have to be integrated out.
These terms, after the use of the BFM as done above, would
bring several corrections 
to the argument of the  determinant (\ref{Seff-det})
 of the following type:
\begin{eqnarray}
\Delta S^{-1} (x-y) = 
\lambda_\alpha
\tilde{S}_\alpha + R  
(x-y) g^2 \lambda _\alpha  \bpsi \lambda_\alpha  \psi + ...
\end{eqnarray}
With a saddle point expansion of the quark determinant 
there will have  leading terms 
linear and quadratic $\tilde{S}_\alpha$ and $\tilde{S}_\alpha \tilde{S}_\alpha$.
By integrating out these  non observable
 colorfull auxiliary fields an additional 
determinant 
can be expanded again 
in a large quark mass limit
 for pions and constituent quark currents.
This would yield extra terms
  that could be added  to those terms found below in this work.
These  resulting contributions would  have  additional quark kernels
$S_0(x-y)$ 
or, equivalently, extra factors $1/M^*, 1/{M^*}^2$  
with respect to the terms of the expansion shown below.
Therefore they are of higher order and numerically smaller in the 
large quark mass expansion.
All these non leading  terms however are outside the scope of the present work.
}

There is an  ambiguity to define the pion field
due to   chiral invariance \cite{weinberg67}
and  
it is  convenient  to perform a chiral rotation to investigate only fluctuations
around the ground state without the scalar field as a dynamical degree of freedom.
In the Appendix A  two different choices of the pion field 
are exhibited and they are used respectively 
 in Sections (\ref{sec:expansion-W}) and 
(\ref{sec:expansion-U}) to investigate the quark determinant.
In the first of them, for the chiral invariant terms
 the pion field always show up in terms of a covariant derivative 
  ${\cal D}_\mu \vec{\pi}$. In the second and more usual definition, 
the exponential representation with functions $U, U^\dagger$ is chosen.

\subsection{Gap equation}
\label{sub:gap}

The a.f. were introduced without further considerations about
their dynamics, in particular about their behavior in the ground state.
To account for that,
the saddle point equations for the effective 
action above (\ref{Seff}) yield the usual gap equations.
 By denoting the auxiliary fields $\phi_q=S(x,y)$,$P_i(x,y)$,
$V^\mu_i (x,y)$, $\bar{A}^\mu_i (x,y)$ these equations
can be written as
\begin{eqnarray} \label{gap-eq}
\frac{\partial S_{eff}}{\partial \phi_q} = 0.
\end{eqnarray}
These equations for the NJL model and GCM  have been analyzed in many works
for the vacuum,
under external B,   at finite temperatures or quark densities,
including in the complete form which correspond to Dyson Schwinger equations
in the rainbow ladder approximation.
In the vacuum the only possible non trivial homogeneous 
solution might exist for the scalar  field.
The magnetic field is known to 
produce magnetic catalysis and it increases the resulting effective mass
 \cite{Meff-B,chiral-cond,review-B-general,cond-B}.
In particular in Ref. \cite{PRD-2016} the same gap equation was solved
for zero and weak magnetic field 
by considering an effective (and confining) non perturbative gluon propagator
and some numerical values were exhibited.
With the non zero expected value of the scalar field
due to DChSB, the chiral rotation to the non linear realisations of chiral
symmetry are usually performed and the dynamical field 
correspond to the fluctuations around the ground state
which is described by the chiral condensate.
This happens at the expense of defining a quark effective mass.
The quark determinant (\ref{Seff-det}) will therefore, from here on, be composed by
the quark kernel given by:
\begin{eqnarray}
S_{0,c} (x-y) &\to& 
S_{0,c}^{-1} (x-y) \equiv i \slashed{D}  - M^*,
\end{eqnarray}
where $M^* = m + \bar{S}$ where $m$ is the current quark mass
and $\bar{S}$ the scalar quark condensate.
When a particular  gauge is choosen for $R$ and $\bar{R}$, the gauge 
fixing parameter can be determined by 
a condition of gauge independence such as:
$\frac{\partial S_{eff}}{ \partial \lambda } = 0$.
All
the quantities in the effective action found below for quarks and pions,
 and also the gap equations above,
 depend basically on the gluon propagator and on 
the original QCD Lagrangian parameters: u-d current quark masses, gauge coupling   $g$,
 a gauge fixing parameter  $\lambda$.

To exhibit a numerical solution for the gap equation above, 
let us consider
an  effective  longitudinal (confining)  gluon propagator
from \cite{cornwall,greensite}
form of: 
$g^2 R^{\mu\nu}_{ab} (k) = K_F/(k^2 + M_g^2)^2 g^{\mu\nu} \delta_{ab} $ 
where $K_F= 8\pi^3 M^2/9$, $g^2$ is the bare gauge coupling constant
and $M_G^2$ an effective gluon mass
\cite{gluonmass,SD-rainbow}. 
This effective propagator 
presents the  strength of the  zero momentum
running coupling constant and a qualitative behavior of the deep
infrared behavior.
The gap equation, as well as all the 
expressions for the effective  coupling
constants derived below are finite, i.e. free of ultraviolet and 
infrared divergences.
By considering $M_g = M \simeq 378$ MeV
  \cite{cornwall} and 
 a current quark mass $m=10$MeV,
 it  yields, for $B_0=0$, $s_{0} \simeq 210$ MeV 
for the scalar auxiliary field $s$ as defined,
for which 
$M^* \simeq  220$ MeV.
The
dependence of the chiral condensate, and therefore
the  quark effective mass, 
on the constant magnetic field 
 has been investigated extensively 
 \cite{review-B-general,ritus,bali-etal1303.1328}.
The increase of the density of states  by accounting 
the lowest Landau levels with high degeneracy in this regime
yields the increase of the chiral condensate.
This so called magnetic catalysis 
has also been related to the positivity
of the scalar QED $\beta-$ function \cite{cond-B}.
The resulting effective quark mass  dependence on the weak magnetic field was analysed in 
\cite{PRD-2016}.

\section{ Large effective masses expansion for the  Weinberg   pion field}
\label{sec:expansion-W}

By performing the first chiral rotation as presented in the Appendix
A,
in expressions (\ref{quark-D}-\ref{q-p-W-L}) for the so called 
Weinberg pion field definition,
the quark determinant can be written 
 (by omitting the dependence on spacetime coordinates)
as:
\begin{eqnarray}   \label{exp-1}
S_{d} &=& C_0  +
\frac{i}{2} \;
Tr \ln
\left\{
\left(    
1 + 
{S}_{0,c}
\left[
 \Phi_W + 
  \left(\sum_q \bar{a}_q \Gamma_q j_q \right)
\right]
\right)^*
\right.
\nonumber
\\
&& \left.
\left( 
1 + 
S_{0,c} 
\left[ \Phi_W  
+  
 \left(\sum_q \bar{a}_q \Gamma_q j_q \right)
\right]
\right)
\right\},
\end{eqnarray}
where 
 the following quantities were defined:
\begin{eqnarray} \label{pi-weinberg}
 \Phi_W  = \Xi &=&
\gamma^\mu \vec{\sigma} \cdot {\cal D}_\mu \vec{\pi} i \gamma_5 
+ 
 i  \gamma^\mu \vec\sigma \cdot 
\frac{\vec{\pi} \times \partial_\mu \vec{\pi}}{1+{\vec{\pi}}^2} 
+  4  m  
\left( 
\frac{\vec\pi^2}{1+ \vec{\pi}^2}
-  \frac{\epsilon_{ijk} \sigma_k \pi_i \pi_j }{1+ \vec{\pi}^2}
\right)
,
\\ \label{bilinears}
 \sum_q  a_q \Gamma_q j_q (x,y) &=&
-
K_0  \bar{R}^{\mu\nu} (x-y) \gamma_\mu  \sigma_i \left[
 (\bpsi (y) \gamma_\nu  \sigma_i \psi(x))
+   i \gamma_5   (\bpsi  (y)
i \gamma_5 \gamma_\nu  \sigma_i \psi (x)) \right]
\nonumber
\\
&+& 
 2  K_0   R (x-y) 
 \left[  (\bpsi (y) \psi(x))
+
 i  \gamma_5 \sigma_i  (\bpsi (y) i \gamma_5  \sigma_i \psi (x)) \right],
\\
C_0 &=& 
\frac{i}{2}  \; Tr \; \ln \bar{S}_{0,c}^{ -1} S_{0,c}, 
\end{eqnarray}
where $K_0 = \alpha g^2$ and 
 $C_0$ contains
corrections to the electromagnetic field effective action that will not
be presented here and it reduces  to 
a constant in the generating functional in the absence of the background photon
 field.
The leading quark-pion and pion
effective interactions in the absence of magnetic field
correspond to the Weinberg large Nc  EFT  and they
have already been obtained  in \cite{EPJA-2016}.
Higher order terms of the expansion generate non leading 
contributions for the pion-quark couplings, however they can be seen to be
suppressed by higher order powers of the kernel $S_0$
in the large quark mass limit.
{ 
By neglecting the  constituent quark currents 
this determinant has been investigated  mainly for
the large quark mass regime, or  weak pion field,
and more generally weak light mesons fields 
without the electromagnetic couplings in different works.
 The resulting leading terms  shown below are written with 
 coefficients (effective coupling constants) 
which contain momentum integrals 
with  progressively higher  powers of 
 the quark and gluon kernels $S_0(k)$ and $R(k)$.
The internal momentum structure of these terms,
in the zero momentum exchange limit,
 can be written in general as
 $\int_k S_0^n (k) R^m(k)$ for $n=1,2,3..$ and 
$m=0,1,2...$ as presented in the next sections.
Pion normalization constant is $F \simeq 92$MeV that is 
relatively small when compared to 
$M^* \simeq 200-300$MeV and therefore this expansion is compatible with 
the weak pion field expansion.
These coefficients or effective coupling constants contain therefore 
momentum integrals 
$\int_k S_0^n (k) R^m(k)$
will be therefore  progressively numerically  smaller, 
already for  quark and gluon effective  masses of the order of 
200-500 MeV. 
Therefore this action will be expanded for large quark and gluon effective masses
and the strict 
convergence of this expansion will not be addressed in the present work.
}

\subsection{ First order terms}

The expansion of the determinant yields non-local 
interaction terms and therefore form factors 
such as the following one:
\begin{eqnarray}
I_2^p &=& 
i^3
d_1 
2   ( \alpha   g^2)
\; Tr \;  \left[ (( {S}_0 (x-z) S_0 (z-y)
{\bar{R}^{\nu\rho}} (x-y) )) \right. 
\nonumber
\\
&&  \left. 
\gamma_\mu \gamma_5  \sigma_i \gamma_\nu \gamma_5
\sigma_j
\epsilon_{imn} 
\frac{
{\pi}_m  (z)  \partial^{\mu} {\pi}_n (z)
}{
1 + \vec{\pi}^2 (z) } 
  (\bpsi (y) \gamma_\rho  \sigma_j \psi(x) )
\right] ,
\end{eqnarray}
where
$d_n =   i \frac{(-1)^{n+1} }{2 n}$,
 the trace $Tr = tr_C \; tr_F \; tr_D \; tr_{k}$ is composed by 
the traces in all internal quantum numbers, 
color, flavor and Dirac indices, and spacetime with eventual
 momenta  integration.
However in the low energy limit one might resolve the traces above by considering 
zero momentum transfer  between  pions and constituent quarks. 
Therefore,
by  resolving the
effective coupling constants  in the corresponding
 local limit, the non zero first order leading 
effective quark-pion couplings and pameteres
are given by:
\begin{eqnarray} \label{L-1-W-1}
{\cal L}_W^{(1)} &=&
g_{\chi 1}  
{\cal D}^\mu \vec{\pi}
{\cal D}_\mu \vec{\pi}
+  i
2 g_{\pi v}
\frac{ \vec{\pi} \times    \partial^{\nu} \vec{\pi}}{
1 + \vec{\pi}^2 } 
\cdot  (\bpsi  \gamma_\nu  \vec\sigma \psi)
+ i 2 g_{\pi v}
  {\cal D}^\nu {\pi}^i
  \bpsi 
i \gamma_5 \gamma_\nu  \sigma_i \psi ,
\\
\label{L-sb-1-W-1}
{\cal L}_{sb,1}&=&
 g_{sb1}     (\bpsi \psi)
+
g_{\beta sb1} F
 \left( \frac{\vec{\pi}^2}{1 + \vec{\pi}^2}
\right)
  (\bpsi \psi)
-
g_{\beta psb} F^2
\frac{\vec\pi^2}{1 + \vec{\pi}^2}
+ 
g_{\chi sb}  F^4
\left( \frac{\vec\pi^2}{1+ \vec{\pi}^2} \right)^2
,
\end{eqnarray}
where $F$ is the pion decay constant suitably
introduced to provide the correct canonical definition of
the pion field.
In these expressions there are  
chiral invariant and symmetry breaking terms.
The following effective coupling constants have been defined:
\begin{eqnarray} \label{chi1}
g_{\chi 1} 
  &=&
-   i N_c  d_1   8
\; Tr' \;  (( \tilde{S}_2 (k)  ))
,
\\  \label{beta1}
g_{\pi v}
 &=& 
i  
N_c  d_1 
4   ( \alpha   g^2)
\; Tr' \;   (( \tilde{S}_2   (k)
\bar{\bar{R}} (k) )) ,
\\  \label{sb1}
g_{sb1} 
 &=& i 
N_c
d_1  16 M^* (\alpha g^2)  
\; Tr' \;  ((  \tilde{S}_0 (k)
R (k) ))
,
\\  \label{beta-sb-1}
g_{\beta sb1} &=&
 i
N_c
 d_1  64  \frac{m }{F^2} (\alpha g^2)
 \; Tr' \; ((  \tilde{S}_2 (k)
{R} (k) ))
\\ \label{Mpi2-W}
g_{\beta psb} 
&=&
- i 
N_c
d_1 64
 \frac{M^* m}{F^2}   \; Tr' \;  (( \tilde{S}_0 (k) ))
,
\\ \label{chisb} 
g_{\chi sb}  
&=&
 i  N_c
 d_1  128 \frac{ m^2}{F^4} \; Tr' \; ((   \tilde{S}_2 (k)  ))
,
\end{eqnarray} 
where
the symbols $ Tr' \; (( S_m (k) )) $ indicates integration in momentum for
combinations of  the gluon kernel,
$R(k)$ was written in expression (\ref{Rk}),  and the following functions 
for the local limit of the couplings:
\begin{eqnarray} 
\label{ts0}\
\tilde{S}_0 (k) 
&=& \frac{1}{k^2 - M^2} ,
\\ 
\label{ts2}
\tilde{S}_2 (k)  &=& 
\frac{k^2 + M^2}{(k^2 - M^2)^2} ,
\\
\label{bbR}
\bar{\bar{R}} (k)    &=& g_{\mu\nu} \bar{R}^{\mu\nu} (k),
\end{eqnarray}

The first term in expression (\ref{L-1-W-1}) $g_{\chi 1}$ reduces to the
usual pion kinetic term with its coefficient the square pion decay constant
$g_{\chi 1} = \frac{F^2_\pi}{2}$ \cite{PRC,EPJA-2016}.
The effective constant $g_{\pi v}$ is the usual pion vector coupling constant 
which turns out to be equal to the axial coupling constant $g_A$.
There are other terms of higher order in the pion field and they are not shown 
being also of higher order in the large $N_c$ limit.
 
The four terms in expression (\ref{L-sb-1-W-1}) are explicit symmetry breaking terms
being their effective coupling constants  proportional   to the 
 effective quark mass ( $M^*$) and to the current quark mass ($m$): 
$g_{sb 1}$ is a correction to the constituent quark effetive mass,
the second term ($g_{\beta sb 1}$)
 corresponds to the sigma term effective coupling of pions to constituent quarks,
and therefore to nucleons,
the third term provides the Lagrangian mass to the pion
$g_{\beta psb} = 
\frac{M_{\pi}^2}{2} .
$  
Expression (\ref{Mpi2-W}) can be rewritten by considering the gap equation 
 for the 
scalar auxiliary field $\bar{S}$ as to yield the Gell Mann Oakes Renner relation
already presented in \cite{EPJA-2016}:
\begin{eqnarray} \label{GOR}
\bar{S} \equiv < \bar{q} q > &=& - \frac{ M_\pi^2 }{m}  F^2,
\end{eqnarray}
where $M_\pi^2$ is proportional to $M^*$ according to expression (\ref{Mpi2-W}).
The last term $g_{\chi sb}$ contains
 the leading pion self interaction due to explicit 
symmetry breaking.

\subsection{ Leading couplings to the background photon}

The leading effective couplings to photons are obtained 
by expanding the quark kernel.
The next leading terms arise also 
from the second order terms of the quark determinant  expansion that 
will not be presented here since they are smaller by a factor proportional to 
${S}_0$ and therefore smaller by  $1/{M^*}^2$.  
The  leading  corrections to the effective terms (\ref{L-1-W-1}) in the very longwavelength 
and zero momentum transfer  are  the following:
\begin{eqnarray} \label{L-1-W-A} 
{\cal L}_W^{(1)} &=&
 i
g_{\beta V}^{ij} 
F^{\mu\nu}
\epsilon_{imn} \frac{{\pi}_m \times    \partial_{\mu} {\pi}_n }{
1 + \vec{\pi}^2 } 
  (\bpsi \gamma_\nu \sigma_j \psi)
\\
&+&
 i  g_{\beta A}^{ij} 
F^{\mu\nu}
  {\cal D}_\mu {\pi}^i
  \bpsi 
i \gamma_5 \gamma_\nu  \sigma_j  \psi
+ g_{F \chi}^{ij}
 F^{\mu\nu}  
{\cal D}_\mu {\pi}_i
{\cal D}_\nu {\pi}_j
,
\nonumber
\end{eqnarray}
where the effective coupling constants are given below.
The coupling $g_{F\chi}$ provides an electromagnetic correction to the pion decay constant,
the terms $g_{\beta V},g_{\beta A}$
provides a  correction to the
(constituent) quark vector and axial  couplings  to pions
respectively.

The   leading  corrections to the leading
 chiral symmetry breaking couplings due to the background  photon field
are of higher order being given by:
\begin{eqnarray}
\label{L-sb-W-A} 
{\cal L}_{sb,1} &=& 
\left(  
g_{\beta sb F}  F^{\mu\nu} F_{\mu\nu} 
+  g_{\beta sb A}  A_\mu A^\mu
\right)
 \frac{\vec{\pi}^2}{1 + \vec{\pi}^2}
\bar{\psi} \psi
\nonumber
\\
&+& 
( g_{sb 1 F} F^{\mu\nu} F_{\mu\nu}  
+ g_{sb1 A} A_\mu A^\mu )
\bar{\psi} \psi
\\
&+&
 \left( 
g_{\beta psb F} 
F^{\mu\nu} F_{\mu\nu}
+ 
g_{\beta psb A} A_\mu A^\mu
  \right)
\frac{\vec\pi^2}{1 + \vec{\pi}^2}
\nonumber
\\
&+& 
\left(
 g_{\chi sb F} F^{\mu\nu} F_{\mu\nu} 
+ 
g_{\chi sb A}  A_\mu A^\mu
\right) 
\left( \frac{\vec\pi^2}{1+ \vec{\pi}^2} \right)^2
,
\nonumber
\end{eqnarray}

In this expression 
$g_{sb1F}$ and $g_{sb1A}$
 provide  electromagnetic correction to the constituent quark mass,
whereas
 $g_{\beta psbA}$ and $g_{\beta psbF}$ for the pion mass.
After taking the traces  
in Dirac and color indices, by 
 separating the trace in flavor indices in the 
first expressions to explicitate its structure,
 the effective coupling constants in expressions
(\ref{L-1-W-A},\ref{L-sb-W-A}) can be written as:
\begin{eqnarray} \label{g-A-W-1}
g_{\beta V}^{ij} = g_{\beta A}^{ij}   &=& 
 i  e N_c  d_1 8 ( \alpha   g^2) \;
tr_F ( \left[ \sigma_i  , Q  \right]  \sigma_j  )
\; Tr' \;   (( \tilde{S}_2  (k)  \tilde{S}_0 (k)
\bar{\bar{R}}  (k) )) ,
\\  
g_{F \chi}^{ij}   
&=&
   i  e    N_c 
d_1  8 \; tr_F 
( \left[ Q  , \sigma_i \right] \sigma_j )
\; Tr' \; 
 (( \tilde{S}_2   (k) \tilde{S}_0  (k)))
,
\\
\label{g-A-W-3}
 g_{\beta sb F} &=&
-   i e^2  N_c  d_1 \frac{80}{9}   m (\alpha g^2) 
 \; Tr' \; ((  \tilde{S}_2  (k) \tilde{S}_0^2  (k)
\bar{\bar{R}} (k) ))
\\
g_{\beta sb A} 
 &=& -  i e^2   N_c
d_1 
 \frac{80}{9}  m  (\alpha g^2)  
\; Tr' \;  ((  \tilde{S}_2  (k) \tilde{S}_0  (k)
R  (k) ))
,
\\
g_{sb1 F} 
 &=& 
-  i e^2  N_c 
 d_1 
  \frac{40}{9}  M^* (\alpha g^2)  
\; Tr' \;  ((  \tilde{S}_0^3  (k)
R   (k) ))
,
\\
g_{sb1 A} 
 &=& 
-  i e^2  N_c 
 d_1 
  \frac{40}{9}  M^* (\alpha g^2)  
\; Tr' \;  ((  \tilde{S}_0^2  (k)
R  (k) ))
,
\\
g_{\beta psb F} &=&
- i e^2  N_c 
 d_1  \frac{160}{9}
 M^* m   \; Tr' \;  (( \tilde{S}_0^3  (k) ))
,
\\
g_{\beta psb A} &=&
- i e^2 
 N_c d_1 \frac{160}{9}   
 M^* m   \; Tr' \;  (( \tilde{S}_0^2 (k) ))
,
\\
g_{\chi sb A}  
&=&
-  i   e^2  N_c
d_1  \frac{128}{9} m^2 \; Tr' \; ((   \tilde{S}_2   (k) \tilde{S}_0 (k) ))
,
\\
\label{g-A-W-10}
g_{\chi sb F}  
&=&
-  i   e^2  N_c
d_1 \frac{128}{9} m^2 \; Tr' \; ((   \tilde{S}_2  (k) \tilde{S}_0^2  (k) ))
.
\end{eqnarray} 
Electromagnetic corrections to the chiral symmetry breaking 
 terms shown  in expression (\ref{L-sb-W-A}) 
also are all  proportional to the current 
quark mass or effective  valence quark 
mass.
It can be noted  that
 while the leading corrections to the chiral invariant 
effective  couplings in the weak magnetic field regime correspond to 
the dipole  magnetic couplings,
being proportional to $F^{\mu\nu}$,
  the leading correctons to intrinsic
symmetry breaking interactions
are of higher order ($F_{\mu\nu}^2$ or $A_\mu^2$)
and therefore  
they increase slower for weak magnetic field.
Of course the interactions with the electromagnetic field
 break chiral and isospin symmetry.

In Figure 1, the diagrams corresponding to the expressions 
(\ref{L-1-W-A})
for the  weak electromanetic field coupling to the quark-pion 
effective interactions
are exhibited.
The dressed (non perturbative) gluon propagator is indicated
by a wavy line with a full circle
and pion represented by dashed lines. 
There are one  photon (dotted line) 
couplings to the internal quark lines
coupled to  one and two pions 
 in diagrams (1a) and (1b).
The one pion coupling is the axial coupling and the two-pion coupling 
to quarks is the vector one.
In diagrams (1c) and (1d) the corrections to the constituent quark mass are  indicated.

\begin{figure}[ht!]
\centering
\includegraphics[width=110mm]{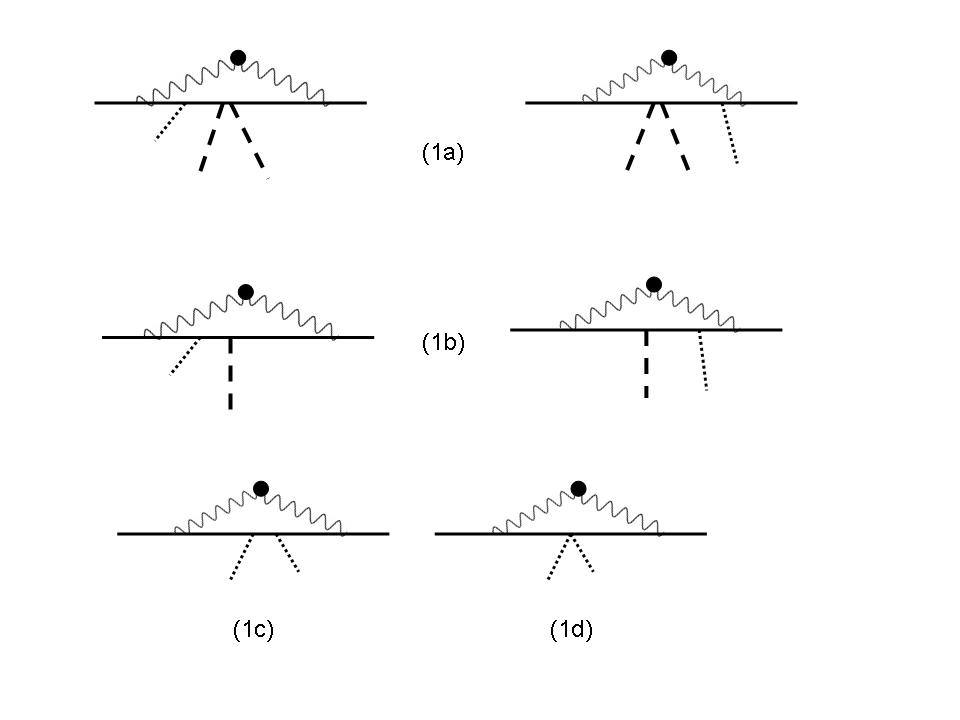}
\caption{ \label{diagrams-W-1}
\small
Diagrams (1a) and (1b) correspond  to the quark-pion effective couplings
from expression (\ref{L-1-W-A}).
 The wavy line with a full dot is a (dressed) non perturbative gluon propagator,
and the
dotted and dashed lines represents respectively 
the  background photon and pion fields,
Diagrams (1c) and (1d) represent the quark mass corrections.
}
\end{figure}

\subsection{ Ratios between effective coupling constants under finite magnetic field }
\label{sec:ratios-W}

Consider now a weak magnetic field   
in the Landau gauge by taking
$ A^\mu  = B_0 ( 0, - x, 0, 0)$,
This magnetic field can  be incorporated into 
the effective couplings  shown  in
(\ref{g-A-W-1}-\ref{g-A-W-10}).
The following terms 
with   redefined effective coupling constants   are obtained:
\begin{eqnarray} \label{L-1-W}
{\cal L}_W^{(1)} &=&¨
 i
\bar{g}_{\beta V}
\epsilon_{ij3}
\epsilon_{imn} \frac{{\pi}_m \times    \partial_{x} {\pi}_n }{
1 + \vec{\pi}^2 } 
  (\bpsi  \gamma_2 \sigma_j \psi)
+
 i  \bar{g}_{\beta A} \epsilon_{ij3}
  {\cal D}_x {\pi}^i
  \bpsi  
i \gamma_5 \gamma_2  \sigma_j  \psi 
+ 
(  \bar{g}_{sb 1 F} + \bar{g}_{sb 1 A})
\bpsi \psi
\nonumber
\\
&+&
\bar{g}_{F \chi} \epsilon_{ij3}
{\cal D}_x {\pi}_i
{\cal D}_y {\pi}_j
+ 
\left( \bar{g}_{\beta sb F} 
+  \bar{g}_{\beta sb A}  
\right)
 \frac{\vec{\pi}^2}{1 + \vec{\pi}^2}
\bpsi \psi
+
 \left( 
\bar{g}_{\beta psb F} 
+ 
\bar{g}_{\beta psb A} 
  \right)
\frac{\vec\pi^2}{1 + \vec{\pi}^2}
\nonumber
\\
&+& 
\left(
\bar{g}_{\chi sb F} 
+
\bar{g}_{\chi sb A} 
\right)
\left( \frac{\vec\pi^2}{1+ \vec{\pi}^2} \right)^2
,
\end{eqnarray}
where the effective coupling constants 
are:
\begin{eqnarray} \label{g-A-W-1-b}
\bar{g}_{\beta V}^{ij} = \bar{g}_{\beta A}^{ij}   &=& 
B_0  {g}_{\beta A}
\epsilon_{ij3}
,
\\   \label{F-chi-B}
\bar{g}_{F \chi}^{ij} 
&=&
B_0 {g}_{F \chi}
\epsilon_{ij3}
,
\\
\bar{g}_{\beta sb F} 
 &=& B_0
g_{\beta sb F}
,
\\
g_{\beta sb A} 
 &\simeq&  - i e^2 B_0^2 d_1 
 \frac{80}{9} m (\alpha g^2)  
\; Tr' \;  ((   \frac{\partial^2 }{\partial q_x^2} 
[\tilde{S}_2 (k)  \tilde{S} (k)
R (k) ] ))
,
\\ \label{g-sb1-F}
\bar{g}_{sb1 F}  
 &=& 
  B_0^2 g_{sb 1 F}
,
\\
\bar{g}_{sb1 A}  
 &=& 
- i e^2 B_0^2 d_1 
\frac{80}{9} m (\alpha g^2)  
\; Tr' \;  ((  
 \frac{\partial^2 }{\partial q_x^2} 
[ \tilde{S}_2 (k)  \tilde{S} (k)
R (k) ] ))
,
\\ \label{beta-psb-F}
\bar{g}_{\beta psbF} &=&
B_0^2 g_{\beta psb F}
,
\\
g_{\beta psbA} &\simeq&
- i e^2 B_0^2 d_1  
 M^* m  \frac{320}{9} \; Tr' \;  (( 
 \frac{\partial^2 }{\partial q_x^2} 
[ \tilde{S}_0 (k) \tilde{S} (k) ] ))
,
\\
g_{\chi sbA}  
&\simeq&
-  i   e^2 B_0^2 
d_1  
 \frac{128}{9} m^2 \; Tr' \; (( 
 \frac{\partial^2 }{\partial q_x^2} 
[  \tilde{S}_2 (k) \tilde{S} (k) ] ))
,
\\
\label{g-A-W-10-b}
\bar{g}_{\chi sb F}  
&=&
B_0^2 g_{\chi sb F}
.
\end{eqnarray}

Some of the possible gauge invariant 
 ratios between the 
$B$- dependent 
(\ref{g-A-W-1-b}-\ref{g-A-W-10-b}) and $B$- independent 
(\ref{chi1}-\ref{chisb}) 
effective coupling constants
in the limit of very large quark effective mass
are given by:
\begin{eqnarray} \label{ratios-W}
(i) \;\; \frac{\bar{g}_{\beta A} }{ g_{\pi v}} \simeq \frac{e B_0}{{M^*}^2}
,
&\;\;\;\;\;\;\;\;\;\;&
(ii) \;\;
\frac{\bar{g}_{F \chi}}{g_{\chi 1}} \simeq \frac{e B_0}{{M^*}^2},
\nonumber
\\
(iii) \;\;
\frac{\bar{g}_{sb1 F}}{g_{sb 1}} \simeq \frac{5 (e B_0)^2}{9 {M^*}^4} ,
&\;\;\;\;\;\;\;\;\;\;\;&
(iv) 
\;\;
\frac{\bar{g}_{\beta psb F}}{g_{\beta psb}} \simeq \frac{5 (e B_0)^2}{18  {M^*}^4} .
\end{eqnarray}
The first and the second of these ratios indicate respectively
 the relative strength of the
leading B-dependent correction for the axial coupling
and of the pion decay constant with their values in the vacuum.
The pion-constituent quark  vector coupling constant  presents the same 
expression as 
the axial coupling constant as shown above, and therefore the 
corresponding ratio is the same.
They  both yield contributions only
for the charged pions due to the isospin tensor $\epsilon_{ij3}$.
The second and third ratios are the leading B-dependent corrections 
to (chiral symmetry breaking) effective quark and pion masses being of higher order.

\section{ Large quark mass expansion for  the second   pion field - $U,U^\dagger$}
\label{sec:expansion-U}

Next let us consider the second pion field definition in terms of 
exponential functions $U$ and $U^\dagger$ presented in 
 expression (\ref{Linear-NLinear})
of the  Appendix.
The quark determinant can be rewritten as:
\begin{eqnarray}  \label{det-UUd}
I_{det} &=& 
\frac{1}{2} Tr \ln 
\left[  \left( S_{0,c}^{-1} + \Xi + 
\sum_q a_q \Gamma_q j_q \right)
\times \left({S}_{0,c}^{-1} + \Xi^* + 
 \left(\sum_q \bar{a}_q \Gamma_q j_q \right)
 \right)^*
\right] ,
\end{eqnarray}
where constituent quark bilinears were given in expression (\ref{bilinears}),
 being that $\bar{a}_s=-a_s$, $\bar{a}_v=-a_v$ and 
 $\bar{a}_p= a_p$, $\bar{a}_a=a_a$.
By neglecting quark bilinears and pion field 
 this determinant   yields a model of the type of 
 the
celebrated Euler Heisenberg effective action for 
for the electromagnetic field 
\cite{EH1936,cond-B,schwinger51,IZ}.
Below,
the leading  quark-pion effective interactions 
and pion self interactions in the presence of the external
photon field are presented.
Most of the  expressions in the absence of the photon field
have been  presented in \cite{EPJA-2016}.
However the pion coupling to the constituent quarks are exploited further below.

The action of the derivative operator and the
background photon coupling
($ {\slashed{D}}$)
on the terms for the pion field, $\Xi$,  can be suitably written by 
making use of the   following pion covariant derivative:
\begin{eqnarray}
D_\mu  U = \partial_\mu U   + i e A_\mu \left[ \frac{\tau_3}{2} \: , \; U \right]
,
\\
(D_\mu U)^\dagger  = \partial_\mu  U^\dagger -
 i e A_\mu \left[ \frac{\tau_3}{2} \: , \; U^\dagger \right]
,
\end{eqnarray}
where it was used that:
$
\left[ Q \; , \; U \right]  = \left[ \frac{\tau_3}{2} \; , \; U \right]
$.

The leading terms of this expansion will be calculated in the zero order 
derivative expansion.
Besides that, 
 only the leading terms in the pion field and in the pion derivative
will be shown, i.e., terms of higher order in $(\pi_i)^m$  (for $m > 2$ or $4$) and
  $(\partial^n\vec{\pi})$  ($n\geq 2)$
 will be neglected.

\subsection{
Leading and next leading pion and external photon couplings 
}

The leading terms of the  pion sector  can be calculated
 to resolve the corresponding  effective couplings
  constants which turn out to be the low energy coefficients
(lec's).
These lec's are written below in a different way than they were
presented in \cite{EPJA-2016}.
By taking the  traces 
 of  Dirac and color,
the leading terms in the very  longwavelength  limit, 
by accounting the leading term from the expansion of the quark kernel
for the bacground  photon field coupling,
 are given by:
\begin{eqnarray} \label{L-U-1}
{\cal L}_{1}  &=& 
-
g_1 \; tr_F \; (U +  U^\dagger)
+ \frac{F^2}{4} 
g_c 
\; tr_F \; {D}^\dagger U^\dagger 
{D} U 
-
l_5 \; e^2 \; tr_F \left( Q  F_{\mu\nu} U^\dagger Q F^{\mu\nu} U \right) ,
\end{eqnarray}
where $tr_F$ stands for the trace in flavor indices and
\begin{eqnarray} \label{g1}
g_1 &=&  
-  i d_1 4 M F N_c  \; Tr' \; (( \tilde{S}_{0} (k)  )),
\\ \label{gc}
g_{c}  g_{\mu\nu}
 &=& 
  i  d_1 16 g_{\mu\nu} N_c \; Tr'  \;
(( \tilde{S}_{0}^2 (k)  )) ,
\\ \label{l5}
l_5 
&=& 
- i  d_1  2 F^2 N_c \; Tr' \; (( \tilde{S}_{2} 
(k) \tilde{S}_0^2 (k) )).
\end{eqnarray}
In expression (\ref{L-U-1}),
the term $g_1$ is the usual leading symmetry breaking term
that yields the pion mass, so that
  it can be written:
$g_1 = \frac{M_{\pi}^2 F^2}{4}$.
With the help of the gap equation (\ref{gap-eq}) 
this term  reduces to
the usual Gell Mann Oakes Renner (GOR) relation
$M  \bar{S}
= M < \bar{q} q > = - F^2 M_{\pi}^2$.
The second term is the lowest order pion kinetic term with the correct
coupling to photons \cite{cpt1,chpt-B}.
The expected
numerical value for the corresponding parameter is $g_c=1$.
The leading correction to the GOR for a weak magnetic field 
will be  presented  below.
The last term $l_5$ is one of the leading chiral symmetry breaking
terms with coupling to the electromagnetic field.

The leading  second order terms are precisely those for 
the next  leading part of chiral perturbation theory.
Again, by resolving effective coupling constants (lec's)
for the  limit of zero  momentum transfer,
they can be written as:
\begin{eqnarray} \label{L-U-4}
I_{\pi- A_\mu} 
&=&
+ \frac{(l_3 + l_4)}{16}  m_\pi^4
 \; tr_F \; ( U^2 + {U^\dagger}^2 ) 
+
\frac{l_4}{8}  m_\pi^2
\; tr_F \;
\left( 
{D}_\mu U^\dagger  {D}^\mu U \right)
\;  ( U + U^\dagger )
\nonumber
\\
&-& 
\frac{l_1}{4}
\; tr_F \; 
(D_\mu U^\dagger D^\mu U)^2
- \frac{l_2}{4}
\; tr_F \; 
 (D_\mu U^\dagger D_\nu U) \cdot
(D^\mu U^\dagger D^\nu U ) ,
\end{eqnarray}
where factors with powers of the pion mass were introduced
 in $l_3,l_4$ coefficients
to compare with usual notation \cite{cpt4}, and 
 the following low energy constants have been defined in the presence of 
an electromagnetic field:
 \begin{eqnarray} \label{l1}
 l_1 
= - \frac{ l_2}{2} 
&=&
  i  d_2  N_c  F^4 64
\; Tr' \; 
(( \tilde{S}_{0,c}^4 (k)   )),
\\  \label{l3-4}
(l_3 + l_4) &=&
-  i d_1   N_c  16 \frac{F^2 M^2}{m_\pi^4}
\; Tr' \;
(( \tilde{S}_{0,c}^2 (k)  )) ,
\\ \label{l4}
l_4 
 &=& 
 i \frac{1}{m_\pi^2}
d_2 N_c 4 \times 12
F^3 M 
\; Tr' \;
(( \tilde{S}_{0,c}^3  (k)  )) ,
\end{eqnarray}
where 
\begin{eqnarray}
\tilde{S}_{0,c} (k) &=&
\frac{
1
}{ ({k}^2  - \Delta_A - {M}^2)}
,
\end{eqnarray}
and $\Delta_A$ given by
$\Delta_A  = e^2  \hat{Q}^2 
A_\mu A^\mu 
+ \frac{i e}{2} \hat{Q}
 \sigma_{\mu\nu} F^{\mu\nu}$.
In the second order  of the determinant expansion 
there are higher order terms of chiral perturbation theory 
that are not presented here.

In the limit of
$\Delta_A = 0$
 for the expressions above
the effective couplings  (\ref{L-U-4})
 reduce basically to the fourth order terms ChPT weakly coupled to the electromagnetic field
with coefficients defined differently.
By considering the full expressions for the lec's dependence on the 
electromagnetic field there appear corrections for 
the case of strong electromagnetic field, i.e. 
electromagnetic field  dependent low energy constants.
These corrections due to electromagnetic field can also be written 
in the usual way  by expanding $\tilde{S}_{0,c}$
for weak electromagnetic field and this second procedure yields
higher order terms of $B_0$-dependent  ChPT.

There is another leading 
electromagnetic coupling to pions that arise from expanding the 
  the low energy coefficient  $g_c$ for weak electromagnetic field
(\ref{gc}). 
By resolving
the effective coupling by taking the traces in color and Dirac indices,
 it can be written as
\begin{eqnarray}  \label{L-D-F}
{\cal L}_{D - F} &=& 
- i  \frac{l_6}{2} tr_F 
\left[ 
\left(
Q  F_{\mu\nu} (D^\mu U)^\dagger (D^\nu U) \right)
+ 
\left(
Q  F_{\mu\nu} (D^\mu U)  (D^\nu U)^\dagger \right)
\right],
\end{eqnarray}
where 
\begin{eqnarray} \label{l6}
l_6 &=& 
i  d_1 e 4 N_c F^2 \; Tr' \; ((\tilde{S}_0^3 (k) )).
\end{eqnarray}
This completes the  leading terms of ChPT in the presence of a 
weak electromagnetic field  \cite{chpt-B,cpt1}.
{ 
This structureless  pion limit  corresponds to a truncated version of 
more general calculations \cite{PRC,wang-etal}
and the resulting leading terms of Chiral Perturbation Theory
 are the correct ones.
}

\subsection{First order constituent quark-pion effective couplings}

The leading effective constituent quark-pion  terms   arise in the
very longwavelength  limit
are the following:
\begin{eqnarray} \label{pion-q-couplings}
{\cal L}_{q-\pi} &=&
 M_3 \; 
j_s  +
g_{2 js } F \;  tr_F \; Z_+  j_s + g_{ps} F  \; tr_F (\sigma_i Z_-) j_{ps}^i
+  i 2 g_{V}  \;  tr_F ( \sigma_i \partial_\mu Z_+ )  j^{V,\mu}_i 
\nonumber
\\
&+&
 i 2  g_{A}   \;  tr_F ( \sigma_i  \partial_\mu Z_- ) j_{A,\mu}^i ,
\end{eqnarray}
where 
$Z_\pm = \frac{1}{2} (U \pm U^\dagger)$ 
and where
the following effective coupling constants were defined
by calculating the traces in Dirac and color indices:
\begin{eqnarray} \label{M3}
\label{js-pi}
M_3 &=&
i  d_1 16 N_c {M^*}  (\alpha g^2)
 \; Tr' \; (( R (k)  \tilde{S}_0 (k)  )),
\\ \label{gps}
  g_{2 js} = - g_{ps} &=& i d_1 32  N_c   (\alpha g^2)
\; Tr' \; (( R (k)  \tilde{S}_2 (k)  ))
,
\\   
\label{gv-U}
g_V = g_A &=&  -  i d_1 8 M^*  F N_c 
 (\alpha g^2)
\; Tr' \; (( \bar{\bar{R}} (k)  \tilde{S}_0^2 (k) ))
,
\end{eqnarray}
where the functions 
$\tilde{S}_0 (k) $ and
$\tilde{S}_2 (k) $  were defined in (\ref{ts0}-\ref{ts2}),
and $R(k)$, $\bar{\bar{R}}(k)$ were defined previously as well.
The first term is a correction to the constituent quark mass, that 
is not necessarily the same as the effective mass from the gap equation
\cite{PRD2014},
 $g_{2js}$ yields  a scalar coupling of two pions to a scalar quark current,
the term $g_{ps}$ is the usual leading 
pseudoscalar pion coupling to quarks,
 $g_{V}$  the two pion coupling
  to a vector quark current
 and the last
one is  the usual  $g_A$ the axial coupling.
For the chiral sector defined by 
 the Weinberg  pion definition 
given in the previous Section, 
 only the last two effective derivative couplings appear.
For the correct
canonical normalization of the pion field
multiplicative  factors $1/F$  
must redefine the effective coupling constants.

\subsection{ Relations among pion-quark effective couplings}

Although  numerical estimations  of these effective coupling constants
are strongly dependent on the gluon propagator
and on the value for the quark-gluon coupling constant,
it is possible to obtain an estimation of  their  relative strength  
by  considering  their ratios in the large quark mass limit.
It is interesting to note that the functions
$R(k)$ and 
 $\bar{\bar{R}} (k)$
obtained from
 the gluon kernels 
are related by
 $\bar{\bar{R}} (k) = 2 R (k)$.
To mantain the standard  dimensionless   coupling constants,
the axial and vector couplings must be 
 divided by $F$ to cope   with the canonical pion definition.
Some ratios between the effective coupling constants are exact and others
can be obtained for the limit of very large quark effective  mass.
By denoting $g_A'$ and $g_V'$ the dimensionless definition of the effective couplings,
 the following exact and approximated 
ratios between the effective coupling constants were obtained:
\begin{eqnarray} \label{ratios-U}
(i) \;\; \frac{g_{ps}}{g_{2js}} &=& -1 ,
, 
\;\;\;\;\;\;\;\;
(ii) \;\; \frac{g_A'}{g_{V}'} = 1 ,
\\   \label{ratios-U-2}
(iii) \;\; 
\frac{g_{ps}}{g_{A}'} &\sim&  \frac{M^*}{ F}
, 
\;\;\;\;\;\;\;\;
(iv) \;\; 
\frac{g_{2js}}{g_{V}'} \sim - \frac{M^*}{F} .
\end{eqnarray}
Note   that the first two ratios, (i) and (ii),
 are  exact and
equal to one. 
The ratios (iii) and (iv) are approximated and valid in the limit of 
large quark masses.
The ratio (iii) 
is the quark-level  Goldberger Treiman relation 
\cite{GT-ori,GT,SWbook}, and the last ratio (iv)
is a scalar-vector coupling relation 
analogous to the Goldberger Treiman relation.
 The  ratios (i) and (ii) are gauge invariant.

\subsection{ Leading quark-pion couplings with external  photon }

The leading constituent quark-pion effective couplings to photons
are obtained from  the expansion of the  kernel $S_{0c}$.
The leading  terms  in the local limit
  are given by:
\begin{eqnarray} \label{q-pi-A}
{\cal L}_{q-\pi-A} &=&
\left( M_{AA} A_\mu^2
+ M_{FF} F_{\mu\nu}F^{\mu\nu} \right) \;  j_s 
+
 g_{vmd}
A_\mu \; 
j^\mu_{i=3}
\nonumber
\\
&+&
\left(
g_{F-js-\pi} \; F  \;  F_{\mu\nu}^2
+ g_{A-js-\pi} \; F \; A_\mu^2 \right)
  tr_F  ( \left\{ Q ,  Z_+  \right\} Q )
 j_s
\\
&+&  i 
\left(
g_{F-ps-\pi} \; F  \;  F_{\mu\nu}^2
+ g_{A-ps-\pi} \; F \; A_\mu^2 \right)
 tr_F \left(
 \left[ Q, Z_- \right]  \left[ Q ,  \sigma_i \right]
\right)
 j_{ps}^i
\nonumber
\\
&+&  i g_{jV A } \; F \; 
 F^{\mu\nu} tr_F ( \left\{ \partial_\mu Z_+ , Q  \right\}
\sigma_i  + \left[ Q ,\ \sigma_i \right] \partial_\mu Z_+ 
 ) 
j_{V,\nu} ^i
\nonumber
\\
&+& i  g_{jA A } \; F \; 
 F^{\mu\nu} tr_F ( [ Q ,  \partial_\mu Z_- ]  \sigma_i 
+ \left\{ Q , \sigma_i \right\} \partial_\mu Z_- )
j_{A,\nu} ^i 
\nonumber
\; + \; {\cal O}(A_\mu^2),
\end{eqnarray}
where, by taking the traces in Dirac and color indices, 
and also flavor indices for the first  terms 
($ M_{3AA}, M_{FF}$ and $g_{vdm}$)
the following effective parameters and coupling
constants have been defined:
\begin{eqnarray}
M_{AA} &=&   
 i   e^2  d_1 
N_c
{M^*} \frac{80}{9}  (\alpha g^2)
\; Tr' \; (( R (k)  \tilde{S}_0^3 (k)  ))
,
\\
M_{FF} &=& 
  i   e^2  d_1 
N_c
{M^*} \frac{160}{9}  (\alpha g^2)
\; Tr' \; (( R (k)  \tilde{S}_0^4 (k)  ))
,
\\
g_{vmd} &=&
i e d_1 
 N_c  8 
 (\alpha g^2)
  \; Tr' \; 
(( \bar{\bar{R}} (k) 
 \tilde{S}_0 (k)  ))  
,
\\
\label{gfjspi}
g_{F-ps-\pi } = - g_{F-js-\pi} &=&
  i e^2 
d_1  N_c 16
 (\alpha g^2)
   \; Tr' \; ((  R (k)  \tilde{S}_2  (k)  \tilde{S}_0^2 (k) 
+  R (k)  \tilde{S}_0^3 (k) ))
,
\\
\label{gfjs}
g_{A-ps-\pi } = - g_{A-js-\pi} &=&
  i e^2 
d_1  N_c 8
 (\alpha g^2)
   \; Tr' \; (( \frac{\partial^2}{\partial q_x^2} 
 ( R (k)  \tilde{S}_2  (k)  \tilde{S}_0 (k) 
+  R (k)  \tilde{S}_0^2 (k) )   ))
,\\
\label{gjva}
 g_{jV A} = g_{jA A} &=&  
  i e  d_1  N_c
  {M^*} 8 
 (\alpha g^2) \; Tr' \; (( \bar{\bar{R}} (k) \tilde{S}_0^3  (k) )) 
,
\end{eqnarray}
The first two terms in expression  (\ref{q-pi-A}) provide 
corrections to the constituent quark mas.
While $M_{AA}$ arises as a first order expansion of the quark kernel
$M_{FF}$  is obtained in the next order of this expansion,
nevertheless both are of the same order in powers of $1/{M^*}$.
The second term is a vector meson dominance term, 
and it will not be investigated further in this work.
The couplings $g_{F-js-\pi}$ and $j_{F-ps-\pi}$ are 
corrections to the scalar and pseudoscalar pion-quark couplings
due to the electromagnetic field 
and $g_{jVA}$ and $j_{jAA}$ are the corresponding corrections to the 
vector and axial pion coupling to quarks.
The leading corrections to the axial and vector effective couplings
are   dipole couplings to the electromagnetic field, whereas the 
scalar and pseudoscalar are of higher order.
The  pion canonical normalization would introduce
further factors $F$ 
in the expressions for the coupling constants above.

The
 one loop diagrams 
corresponding 
 to the leading terms of expression (\ref{q-pi-A}) 
are shown in Figure 2.
There are constituent quark-one pion couplings to one and two photons - (2b) 
(respectively axial and pseudoscalar couplings) -
and constituent quark-two pion coupling to one or two external photons 
(respectively vector   and scalar couplings)
- (2a).
The corrections to the constituent quark mass 
shown in (2c).

\begin{figure}[ht!]
\centering
\includegraphics[width=100mm]{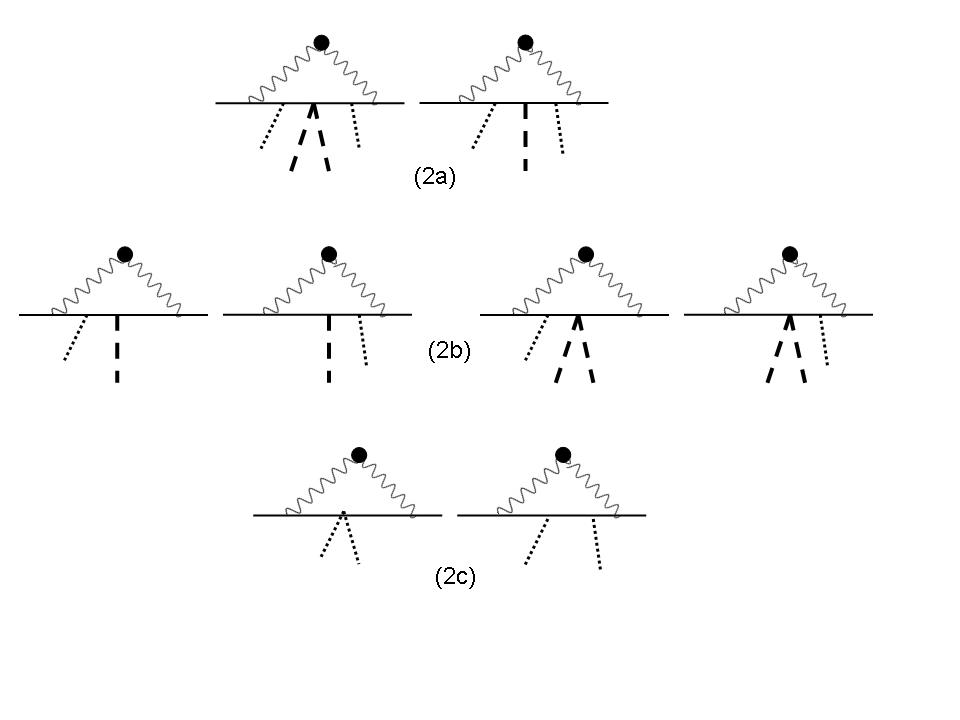}
\caption{ \label{diagrams-2}
\small
In these diagrams, the wavy line
 with a full dot is a (dressed) non perturbative gluon propagator,
a  dotted line represents the external photon field tensor $F^{\mu\nu}$,
 whereas 
the dashed line stands for the pion.
Diagrams (2a) represent the scalar and pseudoscalar pion couplings to quarks
(respectively two pions and one pion couplings)
with further interaction to the photon field. 
Diagrams (2b) represents the axial and vector pion couplings to quarks 
(one and two pion couplings respectively)
with a further  one photon coupling
In diagram (2c) 
the two contributions for the constituent quark effective mass
due to the electromagnetic coupling.
}
\end{figure}

\subsection{ Ratios between effective coupling constants under finite magnetic field }
\label{sec:ratios-U}

Let us  show some approximated ratios in the same limit 
of large quark effective mass
explored in Section (\ref{sec:ratios-W})
by fixing the same magnetic field
$ A^\mu  = B_0 ( 0, - x, 0, 0)$.
From  expression (\ref{q-pi-A}) the following
$B_0$ dependent quark-pion effective couplings can be defined
\begin{eqnarray} \label{q-pi-B}
{\cal L}_{q-\pi-A} &=&
  \bar{M}_{B}   \;  j_s 
+
\bar{g}_{Fjs\pi}   \;
  tr_F  ( \left\{ Q ,  Z_+  \right\} Q )
 j_s
+  i 
\bar{g}_{Fps\pi}  \; 
 tr_F \left(
 \left\{ Q, Z_- \right\} \left\{ Q ,  \sigma_i \right\} 
\right)
 j_{ps}^i
\nonumber
\\
&+& i 
\bar{g}_{jV A }   tr_F ( \left[ \partial_x Z_+ , Q  \right]
\sigma_i 
 ) 
j_{V, y}^i
+  i  \bar{g}_{jA A }  tr_F ( [ Q ,  \partial_x  Z_- ]  \sigma_i ) 
j_{A,y} ^i 
,
\end{eqnarray}
where, by taking the traces in Dirac and color indices, 
and also flavor indices for the first three terms in (\ref{q-pi-A})
($ M_{AA}, M_{FF}, g_{vdm}$)
and $\bar{M}_B$
the following effective coupling
constants have been defined:
\begin{eqnarray} \label{Meff-B-U-2}
\bar{M}_{B} &\simeq& -  i  
 (e B_0)^2  d_1 
N_c
{M^*}  \frac{80}{9}  Tr'  (( R (k)  \tilde{S}_0 (k)  \tilde{S}_x (k)   
+ 3 R (k) \tilde{S}^2_0 (k)  \tilde{S}_x (k) )) ,
\\
\label{gfjspi-B}
\bar{g}_{Fjs\pi} &=&
B_0^2 \; 
g_{F-js-\pi} ,
\\  \label{gps-B}
\label{gfpspi-B}
\bar{g}_{Fps\pi } &=& 
B_0^2 \; 
g_{F-ps-\pi } ,
\\
\label{gjva-B}
\bar{g}_{jV A}  &=&  B_0  \; 
 {g}_{jV A} ,
\\ \label{gvA-B}
 \bar{g}_{jA A}
&=&  B_0 \;  g_{jA A}
.
\end{eqnarray}

In the limit of large effective quark mass the following 
approximated 
ratios are obtained:
\begin{eqnarray} \label{ratios-U-A}
(i) \;\;\;\;\;\;\; \frac{M_{B}}{M_3} \sim  \frac{5}{3} \frac{ (e B_0)^2}{{M^*}^4}   ,
\;\;\;\;\;\;\; 
 &&
(ii) \;\;\;\; \frac{\bar{g}_{F js \pi}}{g_{2 js}} \sim
 \frac{5}{9} \frac{ (e B_0)^2}{{M^*}^4} ,
\;\;\;\;\;\;\;\;\;\;
\\
(iii) \;\;\;
\frac{\bar{g}_{F ps \pi}}{g_{ps}} \sim
 \frac{10}{9}  \frac{ (e B_0)^2}{{M^*}^4} ,
 \;\;\;\;\;\;\; 
 &&
(iv) \;\;\;\;
\frac{\bar{g}_{jVA}}{g_V} = \frac{\bar{g}_{jAA}}{g_A} \sim
2  \frac{ e B_0}{{M^*}^2} .
\nonumber
\end{eqnarray}
These ratios show approximatedly the  strength of the
corrections for the effective couplings
and effective  mass due to the weak magnetic field.
Ratios between B-dependent effective coupling constants from 
expression (\ref{q-pi-B}) can also be extracted, 
some of them are given by:
\begin{eqnarray} \label{ratios-B-U-a}
(i)
 \;\; \frac{\bar{g}_{F js \pi}}{\bar{g}_{F ps \pi}}
 &\sim& 2 
, 
\;\;\;\;\;\;\;\;\;\;\;\;\;\;\;\;\;\;
(ii) \;\; \frac{\bar{g}_{jAA} }{\bar{g}_{jVA}} \sim 1 ,
\nonumber
\\
(iii) \;\; 
\frac{\bar{g}_{F ps \pi}}{\bar{g}_{jAA}} &\sim&   
\frac{e B_0}{{M^*}^2} 
\frac{{M^*}}{F} ,
\;\;\;\;\;
(iv) \;\; 
\frac{\bar{g}_{F js \pi}}{\bar{g}_{jVA}} \sim 
\frac{e B_0}{{M^*}^2} 
\frac{M^*}{ F} .
\end{eqnarray}
These expressions together with 
expressions (\ref{ratios-U-A}) are $B_0$-dependent 
 corrections to the relations (\ref{ratios-U})
and (\ref{ratios-U-2}).

\section{ Numerical estimates}
\label{sec:numerics}

\begin{table}[ht]
\caption{
\small 
 In the first column the chosen values for the quark effective mass 
are displayed
with the corresponding UV cutoff   $\Lambda_i$ and quark gluon coupling 
constant factor $h_a$.
This factor was chosen to reproduce the value of the pion vector or axial 
coupling constant $g_{\pi v} = 1$.
In the second column the gluon propagator  is indicated:
$D_I$ and $D_{II}$ are  the gluon propagators respectively 
from Refs. \cite{SD-rainbow} and  
Ref. \cite{cornwall}.
From the third to the last  columns, values for the effective coupling constants
from  expressions (\ref{chi1},\ref{beta1},\ref{sb1},\ref{beta-sb-1},\ref{Mpi2-W},\ref{g-A-W-1-b},\ref{F-chi-B},\ref{g-sb1-F},\ref{beta-psb-F}) by considering the Weinberg pion field definition.
Results  from the expressions (\ref{g-A-W-1-b},\ref{F-chi-B},\ref{g-sb1-F},\ref{beta-psb-F})
 divided
by factor $(eB_0)/{M^*}^2$ are displayed, 
being therefore independent of the magnetic field
value.
Values of the  Euclidean cutoffs:
$\Lambda_1 = 0.575$GeV, $\Lambda_2 = 0.433$GeV and $\Lambda_3=0.590$GeV.
In the last line (exp. values) some experimental or expected values that reproduce 
experimental data.
All the effective couplings or parameters that depend on the gluon propagator
were multiplied by the same $h_a$ that corrects the strength of the 
quark gluon coupling constant.
} 
\centering 
\begin{tabular}{c c c c c c c c c c c } 
\hline\hline 
$M^*$  ($\Lambda_i$, $h_a$) & $D_i (k)$ &  $f_\pi$ &      
 $g_{\pi v} h_a$ &
$g_{sb1}  h_a$ & $g_{\beta sb1} h_a$ & $m_\pi$ &
$\frac{\bar{g}_V h_a}{(\frac{eB_0}{{M^*}^2})}$  
&  $\frac{\bar{g}_{F\chi}}{(\frac{eB_0}{{M^*}^2})}$ 
 &
     $\frac{\bar{g}_{sb1F} h_a}{(\frac{eB_0}{{M^*}^2})}$ & 
$\frac{m_{\pi}^B}{(\frac{eB_0}{{M^*}^2})}$ 
\\
GeV  (GeV, -)  & -  &   (MeV)  & 
  &  (MeV) &  & (MeV) &  - &   (MeV)$^2$  &  (MeV) & (MeV)  
\\
\hline
\\ [0.5ex]
0.3  \; ($\Lambda_1$, $\frac{1}{3.4}$) & $D_I$ & 92 & 1  &  1470  & 2.3   & 218 & 
0.8 &  0.1  &  111  & 56 
\\ [0.5ex]
0.3 \; ($\Lambda_1$, $\frac{1}{0.2}$) & $D_{II}$  &   92    &  1   & 3420  &  46 & 218 
&  2.5 &  0.1 & 215 &  56
\\ [0.5ex]
0.18  \; ($\Lambda_2$, $\frac{1}{12.8}$) &  $D_I$  &    93
 & 1  &  291 &  23 &  137   &   0.4   & 0.1  & 13
& 26   
\\ [0.5ex]
0.18 \; ($\Lambda_2$, $\frac{1}{0.7}$)  & $D_{II}$ & 93  &  1 & 836   & 39  & 137 & 0.4 &  0.1
 & 31 & 26 
\\ [0.5ex]
0.07  \; ($\Lambda_3$, $\frac{1}{32}$) &  $D_I$  & 212
 & 1    &  81   &  23   & 
138 &  0.2  &  0.1  & 0.8  &  6.3 
 \\ [0.5ex]
0.07  \; ($\Lambda_3$, $\frac{1}{1.8}$) & $D_{II}$  & 
 212   &  1   & 185  &   46  &  138 & 0.5  & 0.1  & 1.8  & 6.3 
\\
exp. values & - &  92.4 & 1 & 313 &  -  &   140 & - & - & -  & - 
\\[1ex] 
\hline 
\end{tabular}
\label{table:results-2} 
\end{table}


In the Table 1 numerical values for some of the effective coupling constants
and parameters of Section (\ref{sec:expansion-W}) are shown, 
in particular those exhibited in expressions
 (\ref{chi1},\ref{beta1},\ref{sb1},\ref{beta-sb-1},\ref{Mpi2-W},\ref{g-A-W-1-b},\ref{F-chi-B},\ref{g-sb1-F},\ref{beta-psb-F}).
The integrations were carried out by performing an analytical continuation to 
Euclidean momentum space and few of them needed an ultraviolet cutoff.
Two very different gluon propagators were chosen to 
make clear the 
ambiguity of the numerical estimates.
The first of the gluon propagators, $D_{I}(k)$,  is a 
transversal component from Tandy-Maris 
\cite{SD-rainbow}
and  the second one, $D_{II}(k)$, is an effective longitudinal  one  by Cornwall
that produces  dynamical chiral symmetry breaking 
 \cite{cornwall}.
 An ultraviolet cutoff was chosen for   the integrals of 
the effective parameters 
 $f_\pi$ and $m_\pi$, such that the at least one of their
the experimental values 
is reproduced. 
Whenever possible, both $f_\pi$ and $m_\pi$ were fited
with the same UV cutoff.
 Different values for the quark effective mass were considered.
For the gluon propagator it was adopted the convention that
\begin{eqnarray} \label{identify}
g^2 \tilde{R}^{\mu\nu} (k) \; \equiv \;  h_a  D^{\mu\nu}_i (k),
\end{eqnarray}
where $D^{\mu\nu}_i (k)$ for
$i=I, II$ 
and $h_a$ is a constant factor associated with the quark gluon coupling constant,
and the reason for considering it is the following.
All the effective parameters that depend on the gluon propagator
and quark-gluon coupling constant were found to exhibit a systematic
deviation from the known experimental values  by nearly the same
factor, 
whenever  they have
known experimental values.
Therefore in the Table these effective coupling constants
were multiplied  by a factor $h_a$ indicated in the first  column of the Table 
and it was chosen to reproduce the constituent quark-pion axial coupling constant
$g_{\pi v} = 1$ \cite{weinberg-2010}.
It corresponds to a particular choice of the quark gluon (running) coupling constant.
The  values that depend on 
the magnetic field, the effective couplings
  are divided by $(eB_0)/{M^*}^2$ in such a way to 
become independent of the magnetic field.
The resulting values listed in the last four  columns   of the Table
 must
be multiplied by $\frac{(eB_0)}{{M^*}^2}$
to be considered in expressions (\ref{g-A-W-1-b},\ref{F-chi-B},\ref{g-sb1-F},\ref{beta-psb-F}).
The factor  $\frac{(eB_0)}{{M^*}^2} << 1$ or $\frac{(eB_0)}{{M^*}^2} < 1$  is the 
parameter for the weak magnetic field expansion.
Therefore all the corrections induced by the weak magnetic field are small.

The numerical values for the estimates of expressions that do not depend on
the gluon propagator are the following:
$f_\pi, m_\pi, g_{F \chi}$ and $m_\pi^B$.
The effective parameters $f_\pi, m_\pi$ and the corresponding
magnetic field dependent parameters $g_{F \chi}, m_\pi^B$
could be expected to depend stronger in the pion structure which has been
neglected  when compared to the work presented in Ref. \cite{PRC}.
The resulting constituent quark effective mass $g_{sb1}$ might be too high and 
no apparent reason was identified, apart from 
the eventual excessively strong contribution from the 
gluon propagator and quark-gluon coupling.
The quark  effective mass and cutoff that best describe known 
experimental data in the Table
 are therefore $M^* = 180$MeV and $\Lambda = 433$MeV.
For the last four columns there are not experimental values available.

In Table 2
some of the effective  couplings and parameters found in 
Section (\ref{sec:expansion-U}) are exhibited,
namely expressions
 (\ref{g1},\ref{gc},\ref{l1},\ref{l3-4},\ref{l4},\ref{l5},\ref{l6},\ref{M3},\ref{gps},\ref{gv-U})
and also (\ref{Meff-B-U-2},\ref{gps-B},\ref{gjva-B}) .
The same logics  and input parameters used in Table 1   
 were considered.
Again the factor $h_a$ in the identification (\ref{identify}) was chosen to 
produce the axial coupling constant to be $g_v=1$ \cite{weinberg-2010}.
The fact
that pions were considered to  be structureless, differently from  Roberts
and collaborators \cite{PRC}, imposes difficulties to reproducing 
the experimental values for $m_\pi=140$MeV and $g_c=1$.
For the last three columns, where the coefficients
for the magnetic field dependent corrections for effective coupling constants are shown,
 there are not experimental values available.
The resulting values listed in the last four  columns   of the Table
 must
be multiplied by $\left(\frac{(eB_0)}{{M^*}^2}\right)^n$ for $n=1, 2$.
Therefore all the corrections induced by the weak magnetic field are very small.
The definitions of the  lec's as written  in expression (\ref{L-U-4}) were chosen in agreement with 
Anderson \cite{chpt-B}, and the numerical values in the Table 2 were 
taken from Bijnens and Ecker \cite{cpt4}.
The resulting numerical values are the one loop renormalized ones
obtained from the following expression:
\begin{eqnarray}
l_i^r = \frac{\gamma_i}{32 \pi^2} \left( \bar{l}_i + \ln \frac{m_\pi^2}{\mu^2} \right),
\;\;\;\; \mbox{ for $i=1,2,3,4,5,6$}.
\end{eqnarray}
being that all the numerical values of 
 these parameters that reproduce 
experimental data  were given in Refs. \cite{cpt4,cpt1}.
Differently from results of the first pion field definition in 
Table 1, 
results from Table 2 indicate that the best choices for the quark 
effective mass and cutoff that reproduce better known 
 experimental data are  
 $M^*=70$MeV and $\Lambda=590$MeV.
This value of the effecitve quark mass might be identified to a constituent quark mass  for pions.
The values for the parameter $g_{vmd}$ for the strength of the 
vector meson dominance effect  are very small when comparing to the
VMD coupling considered for example in Refs. \cite{vmd1}.

\begin{table}[ht]
\caption{
\small 
 In the first column the following set of 
 values    are displayed 
$M^*$ for a given  $\Lambda_i$, $D_i$  for the gluon propagator, $h_a$ 
is the quark gluon coupling constant factor.
Two gluon propagators
$D_i(k)=I$ or $II$
respectively 
from Refs. \cite{SD-rainbow} and  
Ref. \cite{cornwall}.
The sets of values are:
(1) 0.3 GeV,  I, $\frac{1}{2.6}$,
(2) 0.3 GeV,  II, $\frac{1}{0.3}$,
(3) 0.18 GeV,  I , $\frac{1}{4.1}$,
(4) 0.18 GeV,   II, $\frac{1}{0.5}$,
(5) 0.07 GeV,   I, $\frac{1}{5.9}$
and
(6)
0.07 GeV,  II, $\frac{1}{0.8}$.
Values of the cutoffs: for $M^*=0.3$GeV 
$\Lambda_1 = 0.575$GeV, for 
$M^*=0.18$GeV  $\Lambda_2 = 0.433$GeV  
and for $M^*=0.07$GeV  $\Lambda_3=0.59$GeV.
This factor was chosen to reproduce the value of the pion vector or axial 
coupling constant $g_{ v} h_a = 1$ and it multiplies the gluon propagator.
From the second  to the last  columns, values for the effective coupling constants and parameters
from  expressions (\ref{g1},\ref{gc},\ref{l1},\ref{l3-4},\ref{l4},\ref{l5},\ref{l6},\ref{M3},\ref{gps},\ref{gv-U})
and also (\ref{Meff-B-U-2},\ref{gps-B},\ref{gjva-B}) 
by considering the usual  pion field definition in terms of the functions
$U,U^\dagger$.
Results  from the expressions  (\ref{Meff-B-U-2},\ref{gps-B},\ref{gjva-B}) 
 divided
by factors $(eB_0)/{M^*}^2$ are displayed, 
being therefore independent of the magnetic field
value.  
In the last line (e.v.) some experimental or expected values that reproduce 
experimental data. In this line
the  values of 
lec's are extracted from Ref. \cite{cpt4}
according to the definition written in the text.
} 
\centering 
\begin{tabular}{c c 
c c c c c c c c c c c c c } 
\hline\hline 
sets
&  $m_{\pi}$
  & $g_c$   & $l_1$
  &  $l_3+l_4$ & $l_4$ &  $l_5$ &  $l_6$ & $M_3 h_a$ & $g_{ps} h_a$ & $g_v h_a$ & 
$\frac{\bar{M}_B h_a}{(\frac{eB_0}{{M^*}^2})^2}$ & 
$\frac{\bar{g}_{ps}^B h_a}{(\frac{eB_0}{{M^*}^2})^2}$ & 
$\frac{\bar{g}_V^B h_a}{(\frac{eB_0}{{M^*}^2})}$ & $g_{vmd} h_a$
\\
- &     (MeV)     &  - 
  &   $10^{-4}$ & $10^{-1}$ & $10^{-2}$ & $10^{-4}$  & $10^{-3}$ 
& (MeV)  & -& - & (MeV) &  - &  - & (MeV)$^2$
\\ 
\hline
\\ [0.5ex]
(1)  &  218
 &  0.1   &  0.3  &  1.8 &  1.5 &  2.9 &  1.8 & 
1752 & 2.6    &   1 &  628  &  1.8  & 1.1  & 0.2
\\ [0.5ex]
(2)  &  218  & 0.1  &  0.3  &
1.8 & 1.5 & 2.9 & 1.8  & 1630 & 
3  & 1  & 860  &  2.3  & 1.1 & 2.1
\\ [0.5ex]
(3) &  137
&  0.4 & 2 &  1.1 &  2.0  & 8.3  & 4.9
& 908   &   6.2  & 1  &  169    &  0.8  &  0.8 & 0.2
\\ [0.5ex]
(4)   &   
137 & 0.4  & 2  & 1.1 & 2.0 & 8.3 & 4.9 & 
778 & 5.4 & 1 & 260 & 
1.0 & 0.8 &  2.0
\\ [0.5ex]
 (5) & 
138  &  1.0 & 95   & 0.54  & 6.5 
 &  55 &  33 &   440  &  11  & 1   &  18.5   &  0.2   &  0.6 & 0.1
 \\ [0.5ex]
(6) &   138  & 1.0  & 95 
 & 0.54 & 6.5 & 55 & 33  &
354 &  9  & 1  & 25   &  0.2   & 0.6 & 1.6
 \\ [0.5ex]
e.v. &  140  & 1.0  &  88   &   0.6  & 6.8   &  55  & 31   & 313 
&  13.5  & 1 &  -   &  -   &  - & -
\\[1ex] 
\hline 
\end{tabular}
\label{table:results-2} 
\end{table}

\section{Summary and final remarks}

The leading magnetic field corrections to constituent 
quarks couplings to pions were found
  by considering one loop quark polarization 
for a  dressed  one gluon exchange quark  interaction.
For that, the quark field 
was separated into  sea and constituent components by means of the 
background field method and the leading low energy quark-antiquark excitations
were introduced by means 
of the auxiliary field method as it is usually done being that 
 the lighest one, the pion field, was considered.
The use of auxiliary fields contributes to an improvement of
the one loop background field method by incorporating DChSB and the emergence of the 
scalar quark-antiquark condensate and then of the quark effective mass.
The valence quark determinant was expanded for large quark and gluon
effective masses
by considering two different definitions of the pion field.

Firstly, the  Weinberg's  pion field  definition in terms of covariant pion and quark
derivatives was used  in Section (\ref{sec:expansion-W}). 
The leading terms of the quark determinant  expansion
turn out to be
the constituent chiral quark model  \cite{manohar-georgi}
in the version discussed in \cite{weinberg-2010}
that corresponds to 
 a large $N_c$  EFT \cite{EPJA-2016},
 with the   corrections due to the interaction
with the electromagnetic field. 
As discussed
in Ref. \cite{EPJA-2016}
the relative ambiguity in separting the quark field into
valence and constituent quark fields 
corresponds to the
ambiguity of determining the relative contribution of constituent quarks
sector
and pions (or pion cloud) sector to describe hadron observables
in the constituent chiral quark model \cite{weinberg-2010,kalbermann}.
This issue deserves further investigation.
Effective constituent  quark-pion couplings were derived in the presence of the
background  photon field.
Some ratios  between the effective coupling constants
for the limit of very large quark effective mass and weak magnetic field were found.

Secondly, the more usual pion field representation in terms of the 
exponential functions $U,U^\dagger$ was considered in Section (\ref{sec:expansion-U}).
Concerning the coupling to the external electromagnetic field,
the resulting leading terms in the pion sector correspond to 
the leading terms of Chiral Perturbation Theory
coupled to  photons which were found 
up to the fourth order in perfect agreement to the usual formulation
 \cite{cpt1,cpt4,chpt-B}.
It was shown however that some of the higher order
photon-pion couplings in ChPT might be considered as 
electromagnetic corrections to some of the leading lec's.
Therefore  it might be useful to consider  ChPT calculations for stronger electromagnetic 
 fields   by considering  $B_0$-corrections 
for the values of the lec's.
This pion field 
definition makes possible the emergence of
different known  pion effective couplings to quarks:
vector, axial, pseudoscalar and scalar as shown in expressions
(\ref{pion-q-couplings},\ref{q-pi-A}).
These expressions extend and complete previous work \cite{EPJA-2016}.
The corresponding couplings to the electromagnetic field explicitely
break  chiral and isospin symmetries,
and they 
are relatively small with respect to the 
original pion-quark couplings because
$\frac{(eB_0)}{{M^*}^2} < 1$  or $\frac{(eB_0)}{{M^*}^2} << 1$.
In this quark-pion sector the usual relevant effective couplings
 receive dipolar corrections of the order of 
$F_{\mu\nu}$  (vector and axial effective couplings)
  or higher order ones  of the order of $F_{\mu\nu}^2$ and $A_\mu^2$ 
(scalar and pseudoscalar effective couplings).
For larger magnetic fields
the above  expansion   may be reliable 
 by taking into account 
 higher orders terms terms which 
 produce  effective 
 interactions dependent on $[e B_0/{M^*}^2]^{n}$
for $n=1,2...$.
{ The complete account of the 
Landau orbits   
\cite{review-B-general,ritus} appears  to be equivalent 
to the resulting  series in powers of the magnetic field, such as 
it has been done in this work,
as shown explicitely in Ref.  \cite{weak-B}.}
It is interesting to note that, in the leading order terms, 
 the weak magnetic field does  not
mix the contribution of each of the gluon propagator components,
transversal or longitudinal,
for a particular effective coupling constant or parameter at this level of calculation.
 
Approximated and exact 
ratios between the effective couplings and parameters were 
extracted in the limit of large quark effective mass.
They were exhibited  in expressions
(\ref{ratios-W},\ref{ratios-U},\ref{ratios-U-2},\ref{ratios-U-A},\ref{ratios-B-U-a}),
and they  were found to yield expressions such 
as the emblematic 
Goldberger Treiman and GellMann Oakes Renner relations, besides
 new other relations
 with corresponding leading
corrections due to weak electromagnetic (magnetic) field.
Numerical estimations for the effective coupling constants and 
parameters were  presented by choosing two very different gluon propagators
from references \cite{SD-rainbow} and \cite{cornwall}.
The strength of the quark gluon coupling constant was normalized to 
produce the expected value  for the vector pion coupling to constituent quarks
that is equal to the axial coupling.
{ Pion structure and gluon 3-point and 4-point  Green's functions 
do contribute for the resulting form factors, 
effective coupling constants and parameters 
 \cite{PRC,holdom,wang-etal} and these were not investigated in the 
present work.
}
The magnetic field dependent coupling constants
should be seen as partial contributions to the complete value
because  the gluon propagator and quark-gluon vertex
also can present  magnetic field dependent corrections that might be of 
same order of magnitude of the resulting
 effective coupling constants and 
parameters presented in this work.

\section*{Acknowledgments}

The author thanks  short discussion with 
J. O. Andersen, G.I. Krein
and I.A. Shovkovy.
The author participates of the 
project   INCT-FNA, CNPq-Brazil,  Proc. No. 464898/2014-5.

\section*{Appendix A: Chiral rotations}
\label{sec:chiral-rot}

By freezing the scalar field degree of freedom, chiral invariance yields the 
non linear representation.
There is an ambiguity in defining the pion (and quark and all the fields
respecting chiral symmetry) and different chiral rotations yield different pion field 
definitions \cite{weinberg67,SWbook}.
The quark free  terms and its   coupling   to (scalar and pseudoscalar) chiral fields
are  given by:
\begin{eqnarray} \label{sigmamodel}
 \; \bpsi_2 \left[  i \gamma \cdot  \partial
 -  m  +
 \Phi_L \right] \psi_2
  =  \; \bpsi_2 \left[ i \gamma \cdot  \partial
 - m +
 F ( s + i \gamma_5 \vec\sigma \cdot \vec{p} ) \right]
\psi_2.
\end{eqnarray}
A possible redefinition of the quark and pion fields can be implemented by 
the following transformation
\cite{SWbook,W-pionfield,weinberg67}:
\begin{eqnarray} \label{quark-D}
s =  \frac{1 - \vec{\pi}^2}{1+ \vec{\pi}^2} ,
\hspace{.3cm}
 {p}_i = \frac{2 \pi_i}{1 + \vec{\pi}^2} ,
\hspace{.3cm}
\psi &=& \frac{(1  - i \gamma_5 \vec{\sigma}
 \cdot \vec{\pi})}{\sqrt{1+ \vec{\pi}^2}}  \psi'.
\end{eqnarray}
The above pion- quark coupling 
(\ref{sigmamodel}) can be rewritten as:
\begin{eqnarray} \label{q-p-W-L}
 \bpsi_2' 
\left[
 i \gamma \cdot  \partial
-   m^*
+
   \gamma^\mu \vec\sigma \cdot \left(
\frac{\partial_\mu \vec\pi}{1+ \vec\pi^2} i \gamma_5 +
i 
\frac{\vec{\pi} \times \partial_\mu \vec{\pi}}{1+{\vec{\pi}}^2} 
\right)
+ 4  m  
\left( 
\frac{\vec\pi^2}{1+ \vec{\pi}^2}
-  \frac{\epsilon_{ijk} \sigma_k \pi_i \pi_j }{1+ \vec{\pi}^2}
\right)
\right]
\psi_2',
\end{eqnarray}
where
it was used that 
$ \sigma_i \sigma_j = \delta_{ij} + i \epsilon_{ijk} \sigma_k$.
The last two  terms in this expression correspond to the chiral symmetry breaking couplings
in terms of the current quark mass.
These terms
 yield some of 
the terms proportional to the pion mass or to powers of $\vec\pi^2$ in the resulting
effective model.
The following  notation with  two covariant derivatives is considered:
\begin{eqnarray} \label{chiral-deriv}
{\cal D}_\mu  \vec{\pi} \equiv \frac{\partial_\mu {\vec\pi}}{(1 + \vec{\pi}^2)},
\nonumber
\\
\bpsi_2  \partial_\mu \psi_2 \to
\bpsi_2'   D_{\mu}^{c} \psi_2' \equiv 
\bpsi_2'  \left( \partial_\mu   +  i  \vec{\sigma} \cdot 
\frac{\vec{\pi} \times \partial_\mu \vec{\pi}}{1+{\vec{\pi}}^2} 
\right) \psi_2'.
\end{eqnarray}
The canonically  normalized pion field corresponds to 
$\vec{\pi}'= \vec{\pi} F$
and the pion and valence quark fields will be denoted in the non linear realisation as:
$\pi_i$ and $\psi_2$ respectively.
This redefinition of the fields however induces 
 a non trivial   change in the 
functional measure with terms that do not depend on this pion
covariant derivative. 
These terms  will not be presented in this work.

A different parameterization of
the non linear realization   can be used  for the pseudoscalar 
fluctuations
around the vacuum to rewrite expression
(\ref{sigmamodel}), as discussed in Refs. \cite{PRC,E-R1986},
 by means of :
\begin{eqnarray} \label{Linear-NLinear}
\Phi_L \to \Phi_{NL}
&=&  
F \left( P_R U + P_L U^\dagger \right),
\end{eqnarray}
where  $U = e^{i
\vec{\sigma} \cdot \vec{\pi}}$ 
and 
  $P_{R,L} = (1 \pm \gamma_5 )/2$
are the chirality projectors.
These expressions
allow to rewrite the pion sector
in the standard form  of
 Chiral Perturbation Theory.

\section*{Appendix B: Comparison between pion-quark couplings for the two pion definitions}
\label{sec:pion-int}

For the comparison between the two pion field definitions
for the low energy pion Physics regime, the weak pion field
must be considered.
For the two pion field definitions the following weak 
pion field expansions were considered:
\begin{eqnarray}
{\cal D}_\mu \vec{\pi} &\simeq&
\partial_\mu \vec{\pi}
,\;\;\;\;\;\;\;\;\;\;\;\;\;
 \frac{1}{1 + \vec{\pi}^2} \simeq 1 + ...
\\
U &\simeq& 1 + i \vec{\tau} \cdot \vec{\pi} + ...
,
,\;\;\;\;\;\;
U^\dagger \simeq 1 - i \vec{\tau} \cdot \vec{\pi} +
 ...
\end{eqnarray}

The  traces in flavor  indices of the Pauli matrices with 
the matrix $Q$ given after expression (\ref{Seff}) were computed
for most of the terms 
from expressions (\ref{L-1-W-1},\ref{L-sb-1-W-1},\ref{L-1-W-A},\ref{L-sb-W-A}), 
from the Weinberg pion definition (W),
and 
 for expressions (\ref{pion-q-couplings}) and (\ref{q-pi-A}) from 
the second pion field definition (U).
For the vector pion coupling with the electromagnetic 
coupling it has been used:
$ \epsilon_{ijk} \epsilon_{imn} = \delta_{jm} \delta_{kn} - 
\delta_{jn} \delta_{km}$.
 By using the same   notation for both pion definitions,
that again are written dimensionless $\vec{\pi}$, 
the pion-constituent quark couplings found in this work, 
  with and without  leading
electromagnetic field coupling, can be written 
as:
\begin{eqnarray}
{\cal L}^{q-\pi}_W &=&
2 g_{\pi v}
\epsilon_{ijk} {\pi}_i (    \partial^{\nu} {\pi}_j)
  j_{V,\nu}^k
+  2 g_{\pi v}
 ( {\partial}^\nu {\pi}^i) j_{A,\nu}^i  
\\
&+&
g_{\beta sb1} F
 \vec{\pi}^2
  j_s
,
\\
{\cal L}^{q-\pi}_{W,B} &=&
g_{\beta V} 
T_{jki}  F^{\mu\nu}
{\pi}_j    \partial_{\mu} {\pi}_k
j_{V,\nu}^i
+
  g_{\beta A} \epsilon_{3ij} 
F^{\mu\nu}
  \partial_\mu {\pi}^i
 j_{A,\nu}^j  ,
\\
&+&
\left(  
g_{\beta sb F}  {F^{\mu\nu}}^2
+  g_{\beta sb A}  A_\mu A^\mu
\right)
\vec{\pi}^2
j_s 
\\
{\cal L}^{q-\pi}_U &=&
g_{2 js } F \;  \vec{\pi}^2  j_s 
+ g_{ps} F  {\pi}_i   j_{ps}^i
+  i 2 g_{V}   \epsilon_{ijk} \pi_i (\partial_\mu \pi_j )  j^{V,\mu}_k 
\\
&+&
  2  g_{A}    (\partial^\mu \pi_i)  j_{A,\mu}^i  ,
\\
{\cal L}^{q-\pi}_{U,B} &=&
-
\left( 
 g_{F-js-\pi}   
F_{\mu\nu}^2
+  g_{A-js-\pi}   
A_{\mu} A^\mu
\right) F 
  \frac{5}{9} \vec{\pi}^2  
 j_s
\\
&-&
g_{F-ps-\pi} \; F  \;  F_{\mu\nu}^2
 \frac{4}{3} \epsilon_{ij3} \pi_i
 j_{ps}^j  ,
\\
&-&   g_{jV A } \; F \; 
T_{jki}  F^{\mu\nu} \frac{4}{3}
 \pi_j 
(\partial_\mu \pi_k)
j_{V,\nu}^i
-   g_{jA A } \; F \; 
 F^{\mu\nu}  \frac{4}{3} \epsilon_{ij3}\;  \partial_\mu \pi_i \; 
j_{A,\nu} ^j ,
\end{eqnarray}
where 
$ T_{jki} = \delta_{ij}\delta_{3k} - \delta_{j3} \delta_{ik}$.
The expressions ${\cal L}_{W}^{q-\pi}$ and ${\cal L}_{W,B}^{q-\pi}$
 were extracted from
eqs. (\ref{L-1-W-1},\ref{L-sb-1-W-1},\ref{L-1-W-A},\ref{L-sb-W-A})
and the expressions ${\cal L}_{U}^{q-\pi}$ and ${\cal L}_{U,B}^{q-\pi}$  
were calculated from 
eqs. (\ref{pion-q-couplings},\ref{q-pi-A}).
The     couplings $g_{ps},  g_{F-ps-\pi}$  only 
emerge in the pion definition
in terms of functions $U,U^\dagger$.
For the vector, axial and scalar couplings the following identification 
can be done:
\begin{eqnarray}
g_{\pi v} &\sim& g_V,  \;\; \;\;\;\;\;\;\;\;\;\;\;
g_{\pi v} \sim g_A  ,
\nonumber
\\
g_{\beta sb1} &\sim& g_{2js}, \;\;\;\;\; \;\;\;\;\;\;
g_{\beta V} \sim \frac{4F}{3} g_{jVA} ,
\nonumber
\\
g_{\beta A} &\sim& \frac{4F}{3} g_{jVA},  \;\;\;\;\;\;
g_{\beta sbF} \sim \frac{5F}{9} g_{F-js-\pi},  
\\
g_{\beta sbA} &\sim&  \frac{5F}{9} g_{A-js-\pi} . 
 \end{eqnarray}
Only the free pion terms were presented in both  Sections 3.1 and 4.1
and they are trivially the same by a comparison of expressions 
(\ref{L-1-W-1},\ref{L-sb-1-W-1}) and (\ref{L-U-1}).

 \end{document}